%% file: main.tex
\pdfoutput=1
\documentclass[letterpaper,twocolumn,10pt]{article}
\usepackage{usenix-2020-09}
\usepackage{hyperref}
\usepackage{tikz}
\usepackage{amsmath}
\usepackage{cite}
\usepackage{amsmath,amssymb,amsfonts}
\usepackage{algorithmic}
\usepackage{graphicx}
\usepackage{textcomp}
\usepackage{xcolor}

\usepackage{xurl}

\def\BibTeX{{\rm B\kern-.05em{\sc i\kern-.025em b}\kern-.08em
    T\kern-.1667em\lower.7ex\hbox{E}\kern-.125emX}}

\usepackage{xspace}
\usepackage{makecell}
\usepackage{booktabs}
\usepackage{adjustbox}
\usepackage{amsmath}
\usepackage{multicol}
\usepackage{multirow}
\PassOptionsToPackage{linktocpage}{hyperref}
\usepackage{pifont}
\usepackage[normalem]{ulem}

\usepackage{subcaption}
\usepackage{enumitem}

\usepackage{subcaption}
\usepackage{enumitem}

\newcommand{\footlabel}[2]{%
    \addtocounter{footnote}{1}%
    \footnotetext[\thefootnote]{%
        \addtocounter{footnote}{-1}%
        \refstepcounter{footnote}\label{#1}%
        #2%
    }%
    $^{\ref{#1}}$%
}

\newcommand{\footreference}[1]{%
    $^{\ref{#1}}$%
}

\newcommand{\parhead}[1]{\vspace{0.5pt plus 2pt minus 1pt}\par\noindent\textbf{#1}\hspace{1em plus 0.5em minus 0.5em}}
\usepackage[english]{babel}

\begin{document}

\date{}

\title{\Large \bf You Can't See Me: Physical Removal Attacks on LiDAR-based Autonomous Vehicles Driving Frameworks}

\author{
{\rm Yulong Cao}\\
University of Michigan
\and
{\rm S. Hrushikesh Bhupathiraju}\\
University of Florida
\and
{\rm Pirouz Naghavi}\\
University of Florida
\and
{\rm Takeshi Sugawara}\\
The University of Electro-Communications
\and
{\rm Z. Morley Mao}\\
University of Michigan
\and
{\rm Sara Rampazzi }\\
University of Florida
} 

\maketitle

\input{abstract}

\input{intro}

\input{background}

\input{threat_model}

\input{static_analysis}

\input{eval}

\input{simulation}

\input{tracking_system}

\input{defense}

\input{discussion}

\input{conclusion}

\section*{Acknowledgments}
This research was funded by the JSPS KAKENHI Grant Number 22H00519, NSF under the National AI Institute for Edge Computing Leveraging Next Generation Wireless Networks, Grant Number 2112562, as well as NSF grant CNS-1930041, CMMI-2038215, a gift from Facebook. We thank Zhiyi Chen for the help in physical experiments on the vehicle, and Jennifer Sheldon for the help in proofreading.

\bibliographystyle{plain}
\bibliography{refs}

\input{appendix.tex}

\end{document}

%% file: abstract.tex
\begin{abstract}
Autonomous Vehicles (AVs) increasingly use LiDAR-based object detection systems to perceive other vehicles and pedestrians on the road.
While existing attacks on LiDAR-based autonomous driving architectures focus on lowering the confidence score of AV object detection models to induce obstacle misdetection, our research discovers how to leverage laser-based spoofing techniques to selectively remove the LiDAR point cloud data of genuine obstacles at the sensor level before being used as input to the AV perception.
The ablation of this critical LiDAR information causes autonomous driving obstacle detectors to fail to identify and locate obstacles and, consequently, induces AVs to make dangerous automatic driving decisions.
In this paper, we present a method invisible to the human eye that hides objects and deceives autonomous vehicles' obstacle detectors by exploiting inherent automatic transformation and filtering processes of LiDAR sensor data integrated with autonomous driving frameworks. We call such attacks Physical Removal Attacks (PRA), and we demonstrate their effectiveness against three popular AV obstacle detectors (Apollo, Autoware, PointPillars), and we achieve $45^\circ$ attack capability. We evaluate the attack impact on three fusion models (Frustum-ConvNet, AVOD, and Integrated-Semantic Level Fusion) and the consequences on the driving decision using LGSVL, an industry-grade simulator. In our moving vehicle scenarios, we achieve a 92.7\% success rate removing 90\% of a target obstacle's cloud points. Finally, we demonstrate the attack's success against two popular defenses against spoofing and object hiding attacks and discuss two enhanced defense strategies to mitigate our attack.

\end{abstract}

%% file: intro.tex
\section{Introduction}
Perception systems used in Autonomous Vehicles (AVs) are fundamental elements of autonomy and the foundation of reliable automated decisions for driver safety. These perception systems leverage sensors such as LiDARs, cameras, and radars for obstacle avoidance and navigation control. LiDAR sensors, in particular, are used to capture depth measurements of the vehicle's surroundings with high accuracy in 3D \textit{point clouds} to detect obstacles.  
However, prior research has shown how Autonomous Driving (AD) frameworks are vulnerable to attacks on LiDAR sensors that exploit their perception models, which are used for obstacle detection~\cite{Yan2016CanYT, cao2019adversarial, sun2020towards, xiang2019generating, zhao2020nudge, hau2021object}.

\begin{figure}[t!]
  \centering
    \includegraphics[width=0.45\textwidth]{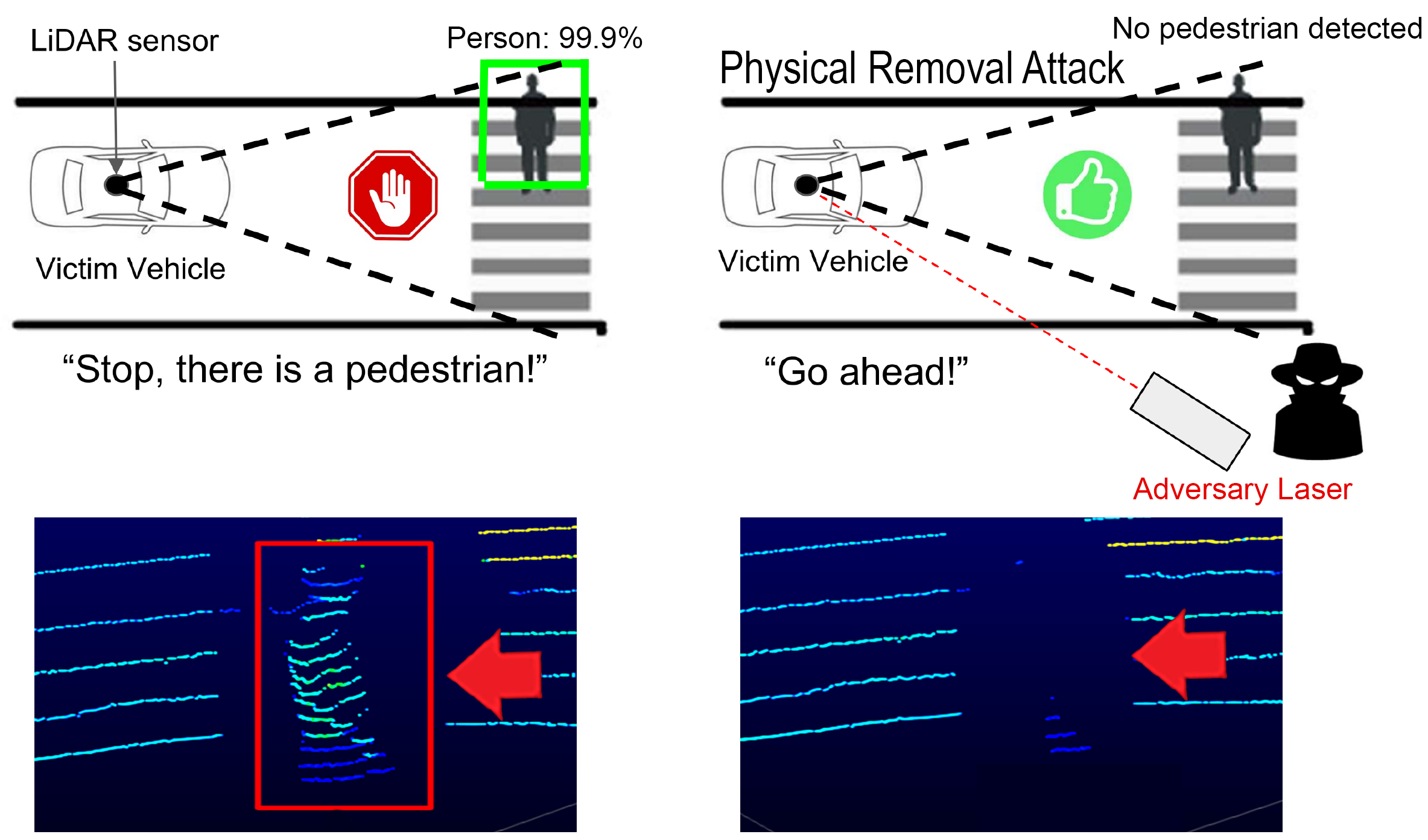}
    \caption{Overview of our LiDAR Physical Removal Attacks (PRA). We exploit the internal automatic filtering of the LiDAR-based perception stack to physically remove selected 3D point clouds from the scene. In this case, the point cloud of a pedestrian.}
    \label{fig:sensor_attack}
\end{figure}

Generally, these attacks focus on creating real-world conditions under which the attacker can manipulate the AVs perception models to "see" non-existent obstacles~\cite{sun2020towards, cao2019adversarial, caoautomated, Shin2017IllusionAD} or to not detect genuine obstacles~\cite{tu2020physically, zhu2021can, Cao2020invisible, hau2021object}. For instance, researchers used laser injections to spoof additional LiDAR cloud points to generate adversarial examples~\cite{cao2019adversarial}, i.e., adding small perturbations obstacles 
to induce misdetection~\cite{hau2021object}.

However, previous works mainly focus on lowering the performance of object detection models, thus it remains unexplored whether there is a physical attack that fully removes LiDAR point clouds from genuine obstacles in real-world driving scenarios and affects the AV driving behaviors.
Thus, we tackle the following research questions in this paper: \textbf{(i)} \textit{Can a real obstacle point cloud be remotely and stealthily removed from LiDAR sensor perception?} \textbf{(ii)} \textit{How can an attacker perform such an attack under realistic conditions?} \textbf{(iii)} \textit{What are the implications of such attacks on AV frameworks and obstacle detection models, and how is it possible to defend against them?}

To answer these research questions, we propose a new attack family, namely \textit{Physical Removal Attacks} (PRA)\footlabel{website}{Video demonstrations, attack traces, and detailed tables of our Physical Removal Attacks can be found at: \href{https://cpseclab.github.io/youcantseeme/}{https://cpseclab.github.io/youcantseeme/}\label{footnote:website}}. 

Leveraging existing laser-based spoofing attacks on LiDAR sensors~\cite{cao2019adversarial}, we investigate the feasibility of manipulating LiDAR sensor data acquisitions to hide genuine obstacles from being detected by AD frameworks, which further raises safety risks for pedestrians and other drivers. We find that, by injecting invisible laser pulses in close proximity to the LiDAR sensor (i.e., below a certain distance threshold), it is possible to force the sensor to discard legitimate cloud points from genuine obstacles in the scene, as depicted in Figure~\ref{fig:sensor_attack}. 
The attack exploits a cascade effect in LiDAR sensors integrated into AD frameworks that rely on two main factors: 1) the intrinsic prioritization of the LiDAR sensors over strong reflections, and 2) the automatic filtering of the cloud points within a certain distance of the LiDAR sensor enclosure.

We first describe the relationship between the internal functioning of the LiDAR sensor and the resulting point removal created by the laser injection. Then we quantify the capability of the attacker in different scenarios. To validate our approach, we examine the effectiveness of the cloud point removal via empirical experiments on a Velodyne VLP-16 LiDAR sensor, where we achieve a $45^\circ$ horizontal attack angle capability. We then analyze the effectiveness of our attack in standard perception systems (e.g., Baidu Apollo~\cite{apollo}, PointPillars~\cite{lang2019pointpillars}, and Autoware~\cite{autoware}), by modeling the attacker's capability to generate a stable spoofing to induce obstacles removal for different scenarios, including different types of obstacles (e.g., cars and pedestrian), at different distances. We further evaluate our PRA on three Camera-LiDAR state-of-the-art fusion models. We demonstrate that the obstacle detection rate drops between 43\% and 76\% in the three tested models, and we are able to cause Autoware Integrated-Semantic Level Fusion~\cite{autoware} failure when the target obstacle is fully removed.  

We then evaluate the attack capability in outdoor scenarios where we achieve the removal of pedestrians and demonstrate the robustness of the attack under different light conditions and at different distances from the spoofer device (up to 10 meters). We also conduct an end-to-end evaluation with an AD simulator~\cite{rong2020lgsvl} to demonstrate the attack consequences of colliding with a pedestrian at a crosswalk or a stopping vehicle by simulating the attack in an ideal setting. Finally, we demonstrate the practicality of attacking moving vehicles (e.g., a robot and a car) with proof-of-concept experiments where we design and prototype a tracking system. We demonstrate a $92.7\%$ success rate in removing $90\%$ of obstacle cloud points with a vehicle driving at 5km/h.

Finally, we systematically investigate existing defenses against laser spoofing attacks and object hiding attacks, demonstrating how they are not effective against our removal attack, and we summarize two defense strategies that we call \textit{Fake Shadow Detection} and \textit{Azimuth-based Detection}. We further discuss and evaluate the defenses on synthesized and real-world traces of our attack achieving 82.5\% True Negative Rate (TNR) and 91.2\% True Positive Rate (TPR) for the first and 99.98\% TNR and 100\% TPR for the second defense.
To summarize, this work aims to model, measure, and demonstrate the capability of removing LiDAR sensor information by leveraging laser-based spoofing techniques, and help defend against the threat to current and future AD frameworks and AVs.

In summary, we highlight the following contributions:
\begin{itemize}[leftmargin=1.3em,topsep=1pt,noitemsep]
    \item We identify and model a laser-based spoofing attack on LiDAR sensors which removes genuine point cloud by exploiting internal cloud point transformations and filtering.
    \item We model the attacker capability, challenges, and limits of PRA on three popular commercial and academic AD perception models (Baidu Apollo~\cite{apollo}, PointPillars~\cite{lang2019pointpillars}, and Autoware~\cite{autoware}). We then evaluate the attack impact on three state-of-art fusion models (Frustum-ConvNet~\cite{fusionFrustum}, AVOD~\cite{fusionAVOD}, and Autoware Integrated-Semantic Level Fusion~\cite{autoware}).
    \item We validate our findings by showing consequences for autonomous vehicles on the production-grade AD simulator LGSVL~\cite{rong2020lgsvl} and conducting real-world attacks on moving robots and vehicles.
    
    \item We verify the effectiveness of the attack against two existing defense approaches for cloud point spoofing: CARLO~\cite{sun2020towards} and hiding attack defense~\cite{hau2022using}. Finally, we propose two enhanced strategies to mitigate the threat.
\end{itemize}

%% file: background.tex
\section{Background and Related Work}

\subsection{LiDAR Sensors in Autonomous Driving}
LiDAR sensors are used to create a map of the car's surroundings, allowing a vehicle to navigate environments.
LiDAR sensors especially are considered more important for AV driving safety decisions over cameras and radars because of LiDAR's higher reliability and accuracy for object detection~\cite{lidar_importance, lidar_intro}.

Popular AD companies such as Google, Uber, Lyft, Baidu, GM Cruise, and Ford, have reached the highest level of self-driving technology, developing AD frameworks integrated with the vehicle controls, whose architecture is based on designs such as Baidu Apollo~\cite{apollo} and Autoware~\cite{autoware, ros_cars}.
Both Baidu Apollo and Autoware integrate LiDAR sensors as well as cameras for obstacle avoidance, and they have been chosen as representative state-of-the-art designs for autonomous transportation technology~\cite{udacity, carma}.
These camera and LiDAR-based AD frameworks usually 
include five main modules: perception, localization, prediction,  planning, and control~\cite{apollo, autoware, garcia2020comprehensive, xu2022sok}. This paper focuses on investigating a vulnerability that affects the perception module used for obstacle detection and avoidance. In particular, our PRA exploit the automated point cloud transformations that can occur in the following components: the LiDAR sensor, the middleware software infrastructure of the AD framework, and the LiDAR data pre-processing phase of the AV perception module. Details of this process will be further discussed in~\S\ref{sec:threat}.

\parhead{3D Spinning LiDAR functioning.}
Among the different LiDAR sensors available, there are two main types commonly used in AVs: spinning and solid-state LiDARs. While spinning LiDARs are a mature technology, solid-state LiDARs are new in the AD context, and they are still under development~\cite{spinning}. Spinning LiDARs use high-grade optics and rotating mechanical hardware, such as a motor, to achieve the full 360$^{\circ}$ view of the vehicle's surroundings.  
Solid-state LiDARs, in comparison, do not have spinning mechanical components, and they usually cover a limited horizontal range ($\sim$120$^{\circ}$).
The present work focuses on 3D spinning LiDARs that are mainly used for AV obstacle detection and avoidance.

Typically, these 3D spinning LiDARs are composed of a vertical stack of Infrared (IR) laser diodes that fire laser pulses along with a corresponding stack of photodiodes. Both stacks rotate, and each photodiode-laser pair covers a particular vertical angle, forming \textit{lines} in the resulting 3D map. If a laser beam is partially or totally intercepted by obstacles in its trajectory, the results of the reflections on the obstacle's surfaces (called \textit{echoes} or \textit{returns}) are backscattered toward the sensor.
The internal timing circuit measures the time between the laser pulse emission and the return at the photodiodes, which is then translated into the distance between the obstacles and the LiDAR. The sensor also captures the
echo's reflectance (intensity of laser pulse returns). By repeating the distance and intensity measurements with a scanning mechanism, the LiDAR covers  360$^{\circ}$  view around the vehicle~\cite{rasshofer2005automotive}.
The resulting collection of these raw signals is called a \textit{point cloud}, where each echo is a cloud point with its own set of X, Y, and Z geometric coordinates in the sensor 3D space, and intensity.

\subsection{Spoofing Attacks on LiDAR Sensors}

Petit et al.~\cite{sensor-blackhat15} and Shin et al.~\cite{Shin2017IllusionAD}
investigated real-world spoofing attacks on LiDAR sensors 
without tampering with the vehicle or broadcasting altered packets in the vehicle network. These works demonstrated that an attacker can inject fake point clouds at various distances
by shining a single laser beam composed of fake pulses precisely synchronized to a spinning LiDAR sensor.
Cao~et~al.~\cite{cao2019adversarial} and Sun et al.~\cite{sun2020towards} improved the attack methodology by injecting up to 200 fake points in the LiDAR field of view. 
Cao~et~al. investigated how this methodology can be used to generate point clouds as adversarial examples to introduce near-front fake obstacles in the scene. 
Sun et al. instead exploited patterns of occluded and distant vehicles to achieve the same goal. Both attacks require a precise pulse injection to generate the adversarial examples and knowledge of the victim LiDAR position in real-time to maintain a robust adversarial pattern when the vehicle is moving~\cite{caoautomated}. 
Hallyburton et al.~\cite{HallyburtonFusionSecurity} use the same attack methodology to demonstrate a black box attack on camera-LiDAR fusion.
The aforementioned attacks assume an attacker shines a laser towards a LiDAR, and thus the spoofing attacks are naturally additive, i.e., injecting additional cloud points into the scene to generate a non-existent object.

\parhead{Object Hiding Attacks.}
Recent works showed that adding adversarial point clouds can also evade object recognition in AD frameworks' ML models~\cite{xu2022sok}. These attacks consist of two main methodologies. In the first attack type, the adversary places physical objects on the target vehicle~\cite{tu2020physically, zhu2021can} or the road~\cite{Cao2020invisible}. These physical objects are built with specific shapes and sizes that minimize their confidence score in the victim AV detection models to become undetected. The process has several constraints in terms of practicality and robustness because building such objects require extreme accuracy.
The second attack type leverages the spoofing of additional cloud points to add small perturbations (e.g., noise) to the target obstacles to change the victim detection model output predictions. 
For instance, Hau et al.~\cite{hau2021object, hau2022using} in their Object Removal Attacks (ORA-Random) proposed using the LiDAR spoofing approach developed by Cao et al.~\cite{cao2019adversarial} to simulate the injection of random cloud points around the target object's bounding box detected by the victim's AV detection model. The authors demonstrate that the generated perturbation in the target object location lowers the detection model's confidence score and causes object misdetection.

In contrast to these previous works that focus on lowering the AV detection model confidence score of certain obstacles, our attack directly removes selected cloud points at a lower sensor level before being used as input by the perception module.
The ablation of the point cloud can thus cause failure in the AD framework when detecting potential obstacles and making an AV planning decision. We show that an adversary can mount an attack to remove specific regions of the LiDAR FOV (Field of View) and consequently hide the presence of genuine obstacles in~\S\ref{sec:prem_analysis}. 
Our attack requires neither knowledge of the perception module output nor the generation of adversarial examples dependent on the perception ML model used by the victim AV. The attack also does not require fine-grained spoofing of the cloud points.

\parhead{Physical Attacks on Sensors.} 
 These attacks are performed using spoofing$/$jamming on cameras \cite{Yan2016CanYT, nassi2020phantom}, LiDARs \cite{cao2019adversarial, Shin2017IllusionAD}, MEMS microphones~\cite{lighcommands2020}, radars, and ultrasonic sensors~\cite{Yan2016CanYT}. Spoofing attacks in particular have been shown to create perturbations to cause the AV's ability to detect obstacles or pedestrians to fail, resulting in potential collisions.
With the increase in popularity of AVs, more attacks against their perception sensors have been conducted in an effort to make AVs more robust and prevent critical system failure. In particular, three studies explored the vulnerability of LiDARs to spoofing and jamming attacks \cite{sensor-blackhat15, Shin2017IllusionAD, cao2019adversarial}.
In this work, we build on these prior works to show that the physical removal of point clouds can lead to AD frameworks failure due to the misdetection of obstacles after enough cloud points have been removed by the attacker.

%% file: threat_model.tex
\section{Attack Overview and Threat Model}
\label{sec:threat}
We give an overview of our removal attack, describe the threat model assumptions, and characterize the PRA principles.

\begin{figure}[t!]
    \centering
    \includegraphics[width=\linewidth]{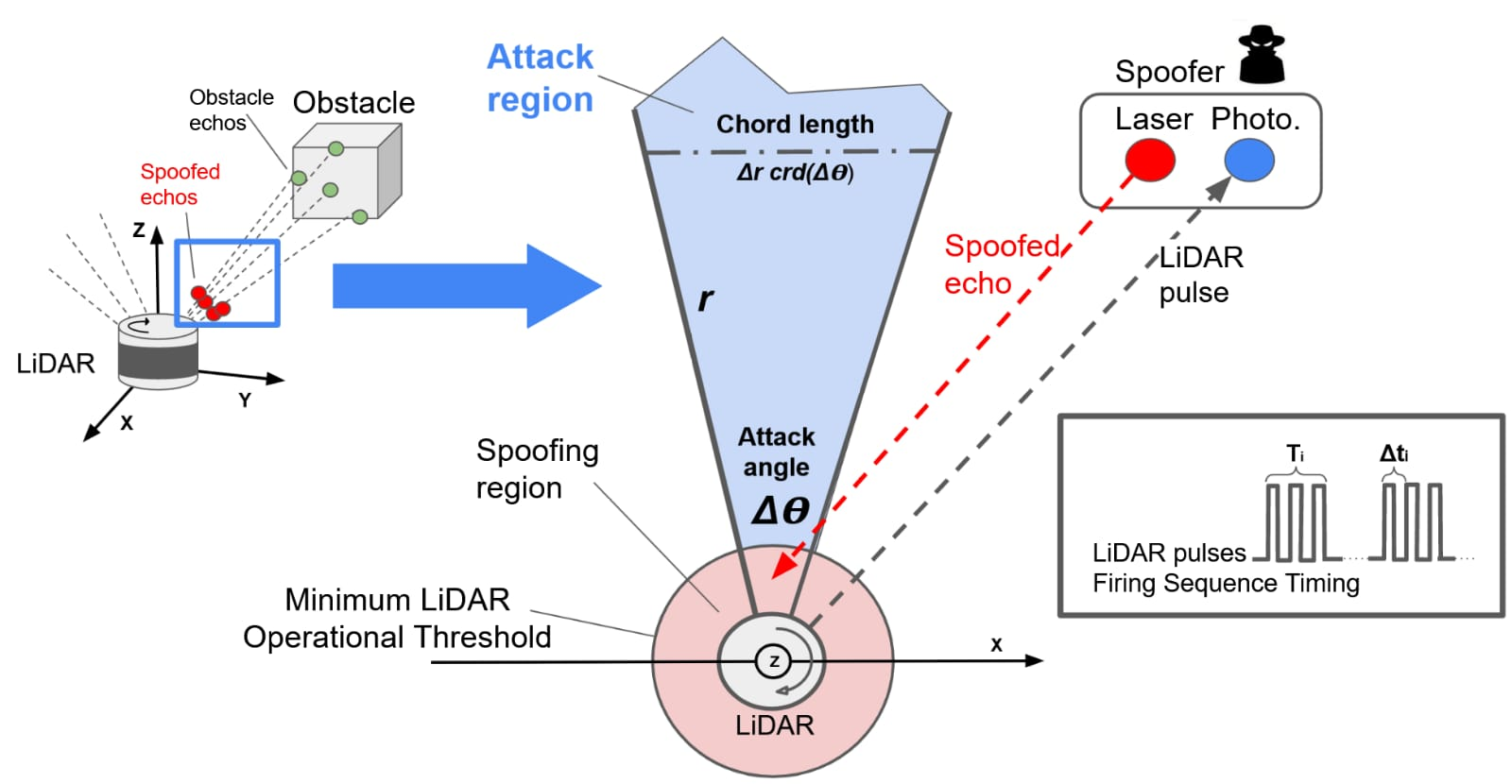}
    \caption{Overview of our Physical Removal Attack. The spoofer photodiode receives the laser pulses from the LiDAR and sends back fake echoes below the LiDAR Minimum Operational Threshold (MOT).}
    \label{fig:optical_diagram}
\end{figure}

\subsection{Physical Removal Attack Overview}
\label{subsec:overview}
Our attack consists of injecting invisible echoes in close proximity of the LiDAR sensor (namely, below a certain distance threshold from the sensor enclosure) in order to force the automatic discard of legitimate cloud points in the scene, such as the cloud points produced by genuine obstacles. The point ablation at the sensor level can, in turn, cause the AD perception module to fail to recognize obstacles and their locations, as we demonstrate in \S\ref{sec:prem_analysis} and \S\ref{subsec:eval}. In other words, genuine cloud point removal is achieved by spoofing cloud points in a specific range between the LiDAR sensor enclosure and the object as illustrated in Figure~\ref{fig:optical_diagram}.  We call this area the \textit{spoofing region}, and its width varies depending on the LiDAR sensor and AD framework used, as described in \S\ref{sec:principles}. We also define the upper limit of this spoofing region, the \textit{Minimum Operational Threshold} (MOT).

To pursue this attack, we show in \S\ref{sec:prem_analysis} how the adversary can use our enhanced methodology based on previous work by Cao et al.~\cite{cao2019adversarial}. Our methodology consists of firing invisible laser pulses that resemble genuine echoes. These pulses are synchronized with the LiDAR sensor firing time sequence to control the location of the spoofed cloud points. As illustrated in Figure~\ref{fig:optical_diagram}, the attacker can synchronize the spoofer device to modify the distance, $\Delta r $, of the spoofed points from the LiDAR where $r$ is the radius of the LiDAR's horizontal FOV. This is achieved by changing the delay of the fired spoofer laser pulses according to the victim LiDAR's laser beam firing sequence timing.
We define \textit{attack angle} $\Delta \theta$ as the LiDAR horizontal angular view affected by the attack, where $\theta$ is the LiDAR horizontal view of 360$^{\circ}$. Note that the attack angle is determined by both the LiDAR model (angle of receiving reflected signals from one direction) and the spoofer (covered region of spoofed signals on the victim LiDAR). Usually, larger attack angles enable the attacker to place the spoofer further away from the target victim and hide objects with a larger view angle.
Knowing the distance, $d$, of a target point cloud to remove (e.g., an obstacle) and the attack angle, we define the resulting \textit{chord length} as $2 d \times sin(\Delta \theta/2)$.

\subsection{Threat Model}

In this paper, we consider the attacker's goal to impact the safety of an autonomously driving vehicle. Our attacker will attempt to accomplish this goal by selectively removing point cloud regions through remote injection of laser pulses in proximity to the LiDAR sensor (below the Minimum Operational Threshold) with a particular timing. 

An attacker can use this strategy to prevent legitimate echoes coming from real-world obstacles in the scene from being perceived by the LiDAR sensor. Consequently, the object detection model of the AV perception module lacks the necessary point cloud information (e.g., distance and location) for detecting potential obstacles present in the scene. The adversary can thus exploit this effect to hide objects, other vehicles, or pedestrians in front of the vehicle for a sufficient amount of time to cause a potential crash or to induce last-second unsafe automatic maneuvering such as steering and drifting into adjacent lanes, thus increasing the risk of being hit by nearby vehicles.

\noindent\textbf{Previous Knowledge.} 
We assume that the adversary can learn the behavior of the victim LiDAR model by reading publicly available documents
(e.g., manuals, datasheets, and open-source code) or by acquiring the same LiDAR used by the victim vehicle.
We also assume that the adversary might only know whether the victim vehicle uses a ROS-based system (e.g., Autoware) or Apollo-based AD frameworks to estimate the spoofing region width of the target AV.
This assumption is less restrictive than prior adversarial attacks on LiDAR and camera perception models that imply white-box attack settings, or having access to the AV detection model output information (e.g., bounding box location) to build adversarial patterns~\cite{cao2019adversarial, Cao2020invisible, hau2021object}. Furthermore, ROS-based and Apollo-based AD frameworks are widely used and considered a standard de facto by a variety of AV companies, including BMW, and Bosch~\cite{ros_cars}.

\begin{table*}[t]
  \caption{Commercial spinning LiDARs: return signal modes available, and default Minimum Operational Thresholds indicated by the sensor manufacturers, ROS middleware, and Baidu Apollo and Autoware AV frameworks.}
  \label{table:min}
\centering
\begin{adjustbox}{max width=\linewidth}
\begin{tabular}{ l | c| c| c | c | c| c }
\toprule
\textbf{LiDAR} & \textbf{Return} & \textbf{Recommended Min.}  &\textbf{Min. Operational} & \textbf{Min. Operational} & \textbf{Min Operational} & \textbf{Literature}\\
\textbf{Model} & \textbf{Modes} & \textbf{Oper. Thr. (cm)}  &\textbf{Thr. ROS driver$^*$ (cm)} &\textbf{Thr. Apollo$^{**}$ (cm)} & \textbf{Thr. Autoware$^{***}$ (cm)} & \textbf{Source}\\
\midrule
 Velodyne VLP-16  & SO$^\bullet$, 2R, L &  100 & 40 & 90 & 40 & \cite{vpl,lambert2020performance, rosdd, apollo}\\
 Velodyne VLP-32c & SO$^\bullet$, 2R, L & 100  & 40 & 40 & 40 & \cite{lambert2020performance, rosdd, apollo}\\
 Velodyne HDL64E-S3 & SO$^\bullet$, 2R, L &  90 & 40 & 90 & 200 & \cite{Vel64, rosdd, apollo}\\
 Velodyne VLS-128 & SO$^\bullet$, 2R, L & NA & 40 & 90 & 90  & \cite{lambert2020performance, rosdd, apollo}\\

 Robosense RS16-RS32-RS128 & SO$^\bullet$, 2R, L & 40 & 20 & 0 & 20 & \cite{robosense, apollo, robosense_ros_driver, autoware}\\

LSLidar C16 & SO$^\bullet$, 2R & 50  & 15 & 30 & 15  & \cite{ros_c16, apollo, manual_c16, yeong2021sensor}\\

 \midrule

 \multicolumn{6}{l}{\footnotesize{SO - Only the strongest return, 2R - Two returns, L - Last return, NA - Not available. ($^\bullet$) Default return mode}.}\\
\multicolumn{6}{l}{\footnotesize{($^*$) Default range for Kinetic and Melodic ROS1 LiDAR drivers. ($^{**}$) Baidu Apollo Version 5.0 and 6.0, including Apollo Cyber RT drivers. ($^{***}$) Autoware.AI Version 1.14.0.}}\\

\end{tabular}
\end{adjustbox}
\end{table*}

\parhead{Spoofing Attack.}
To achieve our PRA, we leverage the laser-based spoofing attacks, which have been proven feasible in previous works~\cite{Shin2017IllusionAD, sun2020towards, sensor-blackhat15, cao2019adversarial, caoautomated}. We modify these attacks to remove genuine cloud points from the LiDAR FOV.  In particular, compared to previous work: i) we improve the Cao et al. setup to achieve better accuracy and performance in the spoofing process, and we increase compactness to improve aiming (as detailed in~\S\ref{sec:prem_analysis}); ii) we assume that the attacker can inject laser pulses at the required intensity and can modify the locations of the spoofed points (distance, altitude, and azimuth) by changing the delay intervals of the spoofer device. However, compared to Cao et al., the attack does not require fine-tuning of the spoofed cloud points locations to generate specific adversarial patterns. The attacker only needs to spoof the points below the MOT of the victim LiDAR sensor. Finally, to aim the laser at moving vehicles, we assume the attacker can use well-known machine learning techniques, such as camera-based object detection and tracking, by collecting LiDAR sensor images from publicly available sources or pre-trained models such as demonstrated in previous work~\cite{caoautomated}.

\parhead{Attack Scenarios.} To perform the attack, we consider several possible scenarios depending on the attacker goal. In this work, we focus on two immediate goals for the attacker: hiding a specific region of the scene (e.g., front-view of the victim vehicle trajectory) and a specific obstacle (e.g., a pedestrian). In both cases, the attacker can place the spoofer device at the roadside, such as close to a pedestrian crossing and intersections as depicted in Figure~\ref{fig:sensor_attack}, to shoot malicious laser pulses at moving AVs. 
If the adversary aims to hide a particular static obstacle, the adversary might also need to know the angular positioning of that object in the front-view of the victim AV to better position the spoofer device. If the obstacle is moving (e.g., a pedestrian), the adversary may use camera-based tracking techniques for detecting the target obstacle position with respect to the moving AV and vary the attack angle and spoofer aiming accordingly. 
Another alternative scenario for hiding a specific region in front of the victim AV consists of placing the attacker device in a vehicle and following the victim car as described in previous works~\cite{cao2019adversarial, caoautomated, Shin2017IllusionAD}.
The laser pulses sent by the attacker are fully invisible to the human eye, and laser shooting devices are relatively small in size, as we will discuss in~\S\ref{sec:prem_analysis}.

\subsection{Attack Principles}
\label{sec:principles}

The basic idea behind PRA is that an adversary can force a LiDAR integrated with the AD framework to automatically ignore genuine cloud points from genuine obstacles under certain conditions. This happens because of two main factors: 1) LiDAR sensors intrinsically prioritize particular echoes (e.g., the strongest) and 2) LiDAR sensors as well as the AV frameworks and middleware automatically filter cloud points closer to the LiDAR sensor enclosure. Both factors are explored below.

\parhead{LiDAR Echo Signals.}
LiDAR sensors, by nature, can receive more than one echo signal from real-world obstacles while firing the laser beams as it widens or diverges over increasing distances.
More specifically, each photodiode in the sensor can acquire multiple echo signals if the laser beam hits multiple objects along its path of propagation. These echoes lie on one line in the 3D space of the LiDAR FOV, and they are typically recorded according to their intensity strength independently from the calculated distance (e.g., strong intensity echoes first)~\cite{man2021multi}. 
However, capturing multiple echoes requires more complex algorithms to elaborate the acquired data, thus spinning LiDAR sensors used in AVs usually capture a limited number of reflections (e.g., up to three echoes), and the default configuration considers only the strongest return (see Table~\ref{table:min}). This is based on the assumption that the echoes with higher intensity come from nearby obstacles that are more critical to detect for AVs~\cite{velodyne64, vpl}.
Capturing multiple echoes (namely recording additional echoes with lower intensity) can give information regarding partially obstructed objects or objects behind semi-transparent surfaces (e.g., a window)~\cite{man2021multi, suarez2005use}. For instance, the Velodyne VLP, HDL, and VLS models support two returns and three configuration modes: only the strongest return (default), the last return received, or both (see Table~\ref{table:min}).

\begin{figure}[t!]
   \begin{minipage}{0.1585\textwidth}
     \centering
     \includegraphics[scale= 0.181]{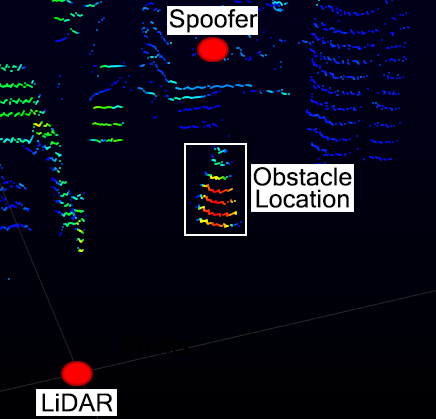}
    \text{(a)}
   \end{minipage}\hfill
   \begin{minipage}{0.1585\textwidth}
     \centering
     \includegraphics[scale= 0.181]{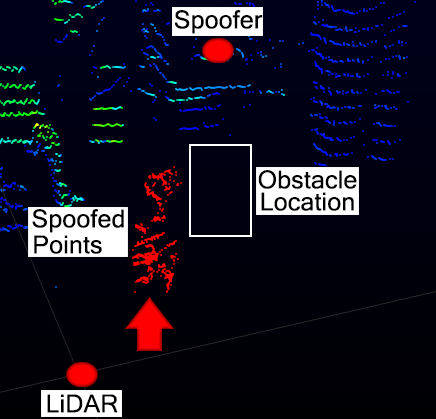}
     \text{(b)}
   \end{minipage}\hfill
   \begin{minipage}{0.1585\textwidth}
     \centering
     \includegraphics[scale= 0.181]{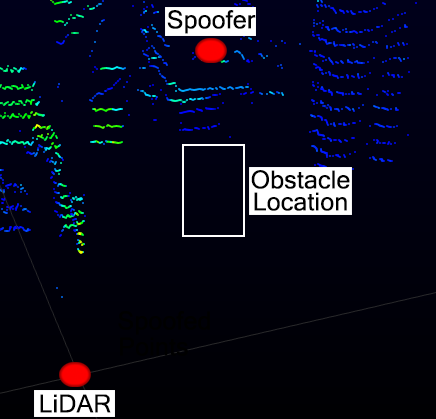}
     \text{(c)}
   \end{minipage}\hfill
    \caption{PRA on a target traffic cone with the LiDAR in dual mode setting. (a) No attack; (b) Spoofed points (in red) between the genuine obstacle and the LiDAR; (c) Spoofed points below the 
    MOT
    of the LiDAR\footreference{website}.}
     \label{fig:ConeCapabilityThree}
\end{figure}

In our attack, we inject high-intensity echoes in close proximity to the LiDAR sensor. This causes the sensor to record our fake reflections as the strongest echoes, ignoring other echoes generated by genuine obstacles farther away. We also demonstrated in \S\ref{sec:prem_analysis} that our methodology allows the attacker to remove cloud points even when the LiDAR sensor is set up to record multiple echoes, as illustrated in Figure~\ref{fig:ConeCapabilityThree}.

\parhead{LiDAR Operational Thresholds.} 

Typically, commercial 3D spinning LiDAR sensors have a minimum horizontal range under which manufacturers do not guarantee full detection of the reflected echoes near the sensor~\cite{minrange1, velodyne64}. 
This range usually varies from 5 cm to 1 m from the sensor enclosure and usually differs from the minimum resolution of the sensor~\cite{vpl, velodyne64}.

In other words, the LiDAR can still receive echoes below this range; however, the resulting cloud point might be inaccurate for certain applications. For this reason, manufacturers recommend a minimum horizontal threshold value,
i.e., MOT,

below which the received echoes should be ignored or discarded (see Table~\ref{table:min}). LiDAR sensors usually implement a preliminary filtering process to automatically discard cloud points below the MOT at the firmware level. This preliminary processing is necessary to normalize the hardware differences due to laser beams and photodiodes alignment, angular resolution of the sensor, rotational speed variation, and to eliminate spurious reflections that might be due to external phenomena (e.g., rain)~\cite{vriesman2020experimental}. For instance, we discover that the LiDAR VLP-16 has an internal MOT of 40 cm from the LiDAR enclosure.

\noindent\textbf{Middleware and AD Framework Automatic Filtering.}
Typically, AD frameworks incorporate a middleware layer to handle the underlying hardware abstraction and the communication between modules, sensors, and actuators. ROS (Robot Operating System)~\cite{ros}, and Apollo Cyber RT~\cite{apollo_cybert} are some of such middleware systems.
In addition to the sensor level filtering, the middleware implements its own filtering process using MOT values that might vary significantly from the LiDAR manufacturer recommendations as illustrated in Table~\ref{table:min}. The filtering at the middleware level is performed usually by ROS device drivers~\cite{rosdd} or custom drivers such as Apollo Cyber RT~\cite{apollo_cybert_drivers}. These drivers discard cloud points outside the horizontal and vertical 3D map
range set in the middleware configuration. 
Finally, the point cloud is filtered again at the AD framework level based on the internal configuration and calibration settings of the specific AD framework (e.g., Autoware or Apollo). For instance, the framework defines the Region of Interest (ROI) and processes the LiDAR point cloud to exclude cloud points outside the ROI. The process usually involves discarding points outside the road, removing background objects (e.g., buildings and trees), and distinguishing ground points to not elaborate from non-ground points. This automatic filtering largely reduces the size and dimensions of the input used by the obstacle detection model to improve framework performance in terms of run-time~\cite{Cao2020invisible}. After the filtering phase, the remaining point cloud is then used for the subsequent pre-processing that generates the input to the machine learning models (e.g., CNN models) used for obstacle detection in the perception module.

\noindent\textbf{PRA Spoofing Region.}
Our attack exploits the presence of the aforementioned automatic cloud point filtering as well as the strong echo prioritization to remove both the injected spoofed points and the point cloud of legitimate obstacles from the LiDAR FOV.
The overall spoofing region width of our removal attack corresponds to the highest MOT value among the sensor, middleware, and AD framework thresholds. For example, with the default settings illustrated in Table~\ref{table:min}, 90 cm is the spoofing region width for the VLP-16 LiDAR integrated with Baidu Apollo, and 40 cm is the spoofing region for the same LiDAR model integrated with Autoware. It is also important to note that, even though the middleware and framework thresholds might be manually set to zero during the AV design, the spoofing region width ultimately depends on the LiDAR sensor's internal MOT, which usually cannot be accessed and modified.

\noindent\textbf{Additional Blind Areas.}
In addition to the automatic filtering process, most of the current AD systems involve placing spinning LiDAR sensors on the top of the vehicle’s roof at a minimum height from the ground to allow a full 360$^{\circ}$ view of the surrounding. This configuration is usually preferred due to the limited vertical FOV of commercial spinning LiDAR sensors (20$^{\circ}$- 40$^{\circ}$ for 64-channel commercial spinning LiDARs~\cite{Vel64, OS2}). 
However, this location can also generate reflections from the vehicle's surfaces (for example due to the reflecting coating of the vehicle's roof) thus AD frameworks usually recommend applying additional cloud point horizontal filters and adjusting the calibration settings of the sensor in the near-field area to remove the spurious echoes from the vehicle while processing the cloud point data~\cite{filtering}.
These discarded areas around the vehicle reduce the overall operational range of the sensor, and inadvertently enlarge the spoofing area suitable for our attack.

%% file: static_analysis.tex
\section{Preliminary Analysis}
\label{sec:prem_analysis}
 
In this section, we investigate the feasibility and capability of the attacker to pursue our removal attack. We first discuss the attack design and evaluate the attack capability in an indoor setting. We then analyze the attack's ability to remove target obstacles. The final part of this section discusses the effect of the attack on the AVs object detection and fusion models.

\subsection{Attack Design}
\label{subsec:attack_design}
To successfully conduct our removal attack, it is important to address the following two challenges: First, how to stably spoof cloud points below the MOT
and, second, how to ensure that the attacker echoes are the ones prioritized by the sensor over the genuine ones. We consider the attack model and spoofer design below, to resolve these challenges.

\parhead{Attack Model.} As illustrated in Figure~\ref{fig:optical_diagram}, the attacker aims to spoof echoes below the LiDAR MOT ($r \leq MOT$) in the spoofing region, so that the genuine echoes are removed. To increase the attack success rate, the attacker  also aims to spoof the largest attack region possible. Hence, the attacker controls the firing timing to match with LiDAR scans on all vertical lines (along the Z axis) and a specific horizontal viewing angle (along azimuth) that we called attack angle $\Delta \theta$. Calculating the precise firing timing 
needs synchronization with the LiDAR. 

Therefore, we use a photodiode to capture signals from the LiDAR per scan so that the spoofer can fire at the correct timings. Finally, since the attack goal of removing an obstacle is dependent upon obstacle sizes and locations, we simplify it as removing an attack region (shown in Figure~\ref{fig:traffic_cone}) that covers the obstacle. Notice that, by controlling the firing timing, the attacker can move the spoofed points at different distances from the LiDAR enclosure with various attack angles as long as it is within the attack capability. Therefore, the attacker can set up a roadside spoofer and remove obstacles at different locations.

\begin{figure}[t!]
    \centering
    \includegraphics[width=0.9\linewidth]{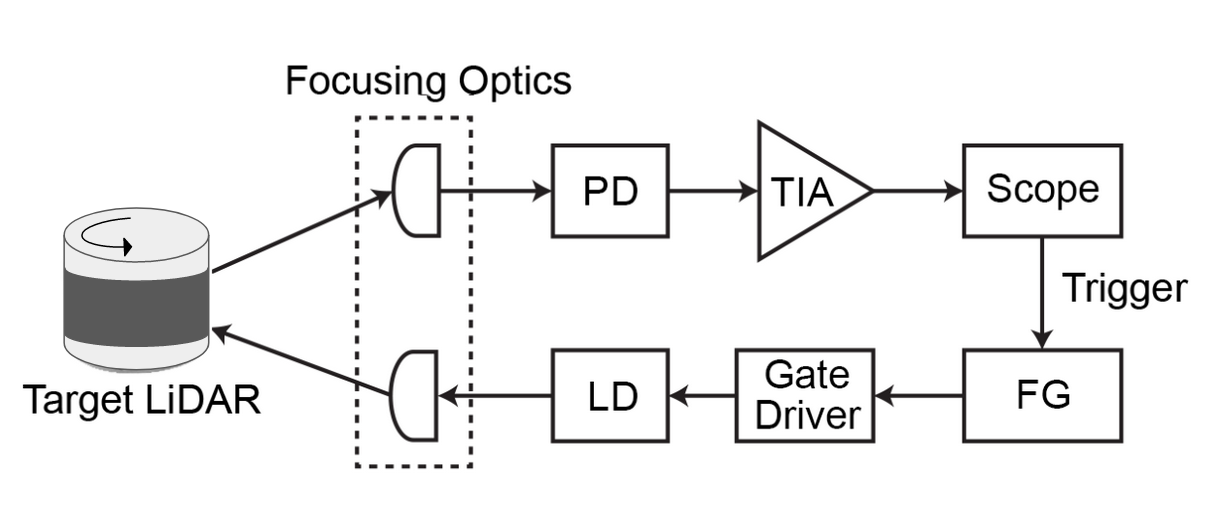}
    \caption{Illustration of the basic hardware components used for the spoofer setup. The laser beams are captured by a photodiode and sent back to the LiDAR sensor with a delay.}
    \label{fig:otpical}
\end{figure}

\parhead{Improved Spoofer Design.} We build the spoofing setup (see 
Figure~\ref{fig:otpical}),
which generates a train of pulses synchronized to the target LiDAR (e.g., Velodyne VLP-16). 

The setup captures a pulse from the LiDAR using
a Hamamatsu S6775 photodiode~\cite{hamamatsu_diode}
and feeds it to an oscilloscope (Tektronix MSO5204) through a simple trans-impedance amplifier (TIA) made by the general-purpose
TL082 operational amplifier. The oscilloscope captures the TIA output with a configured threshold, 
and generates a trigger output synchronized to the LiDAR's scan interval.
Upon receiving the trigger signal, a function generator (Tektronix AFG3201),
operating in the burst mode, synchronously generates a train of voltage pulses. 
Finally, these voltage pulses drive the 
Osram SPL LL90\_3 905-nm laser diode~\cite{osram_diode} through a gate driver~\cite{osram_diode2}.
We use a pair of plano-convex lenses, 1-inch diameter, as collimation and focusing optics.

When the photodiode receives a pulse from the LiDAR, it triggers the pulse firing following the timing given by the function generator. This timing depends on the attacker goal. Since the rotation speed (RPM), and firing sequence timing of the victim LiDAR sensor is constant (and often publicly available from the manufacturer datasheet), the attacker can synchronize the spoofer pulse firing to inject fake echoes in every coordinates of the LiDAR FOV, by changing the timing between the firing.
The basic principle of the above setup follows the previous one by Cao~et~al.~\cite{cao2019adversarial}, 
but is downsized for mounting it on a tracking system to facilitate the aiming that we discuss in~\S\ref{subsec:tracking}. For instance, the S6775 photodiode is much lighter and cheaper than photodiode used in the previous work designs. Similarly, the LL90\_3 laser diode has a built-in driver in its plastic package, eliminating the need for mounting a heavy laser driver on the tracking system.

With the upgraded setup, we are able to spoof more points covering a larger attack region compared to previous works\cite{cao2019adversarial, sun2020towards}: from 200 
to thousands of points as shown in Figure \ref{fig:attack_capability_indoor}. 
The increased horizontal range also opens the space to spoof points at different locations without moving the spoofer as required in the previous works~\cite{cao2019adversarial,sun2020towards}. Therefore, it makes it possible for a roadside attacker to remove cloud points right in front of the victim AV as illustrated in Figure~\ref{fig:sensor_attack}.

\parhead{Multimode Analysis.}
For Velodyne VLP-16, there are three different operating modes: strongest mode, dual mode and last mode where different echoes are used for calculating the point cloud (i.e., strongest mode uses the echo with maximum intensity; last mode uses the last returned echo while the dual mode uses both). We demonstrate that our removal attack removes points for all three modes. As shown in Figure~\ref{fig:ConeCapabilityThree} (c), though the dual mode capture multiple echos, both spoofed echos and echos generated by the genuine obstacle are absent when the injection falls into the spoofing region (below the MOT). This is because our spoofed points' intensity is high enough that they are the only strongest echoes captured by the sensor and automatically filtered out. For the rest of the paper, we use the strongest mode for the experiments if not specified otherwise since it is the typical default mode for AV LiDARs. 

\subsection{Attacker Capability}

While laser spoofing on LiDAR sensors has been validated in the real-world by previous research~\cite{cao2019adversarial}, in
this section, we quantify the capability of the attacker to stably remove 3D cloud points in a selected region.

\label{subsec:attacker_capability}

As described in the attack model, the attacker is already able to spoof all the vertical lines by synchronizing with the LiDAR firing timing, thus, we conducted empirical experiments to investigate the maximum achievable attack angle $\Delta \theta$. We use a Velodyne VLP-16 LiDAR model~\cite{vpl} which is used in previous work~\cite{cao2019adversarial,sun2020towards,Shin2017IllusionAD}. This sensor is composed of a vertical array of 16 laser diodes and corresponding photodiodes to fire laser pulses at different angles following a preset pattern. The VLP-16 has a 360$^{\circ}$ field of view with 30$^{\circ}$ vertical angle range from -15$^{\circ}$ to +15$^{\circ}$, and 0.2$^{\circ}$ of angular resolution. The VLP-16 fires laser pulses in a cycle every 55.296 $\mu$s, with a period of 2.304 $\mu$s. The receiving time window for the reflected echoes is about 667 ns. 
This sensor is currently used in the Baidu Apollo architecture and Autoware; it uses the same design principle as other spinning LiDARs used in autonomous vehicles (e.g., Ouster OS1~\cite{OS1}, Robosense~\cite{robosense}, LSLidar~\cite{manual_c16}) with a similar design and firing pattern.

\begin{figure}[t]
   \centering
   \begin{minipage}{0.46\textwidth}
     \centering
     \includegraphics[scale=0.25]{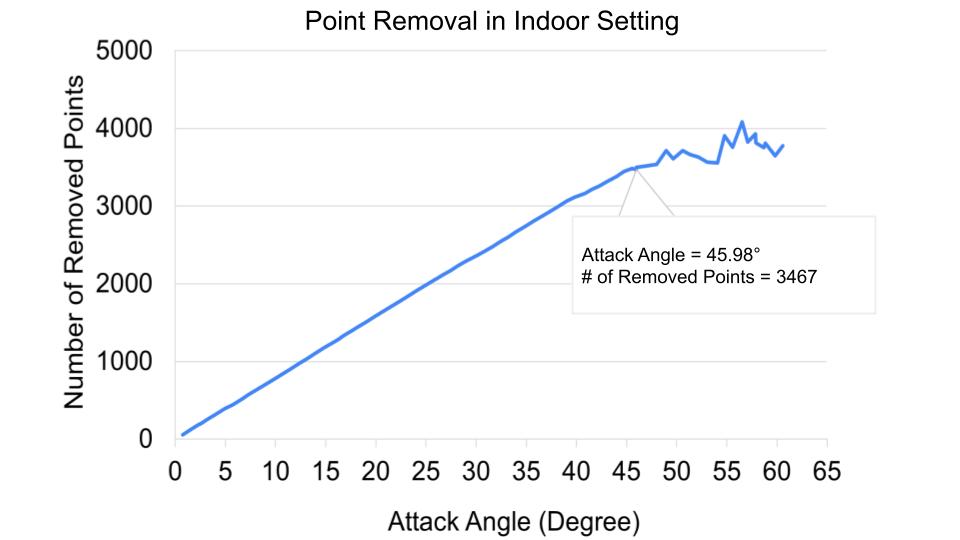}
   \end{minipage}\hfill
    \begin{minipage}{0.46\textwidth}
     \centering
     \includegraphics[scale=0.25]{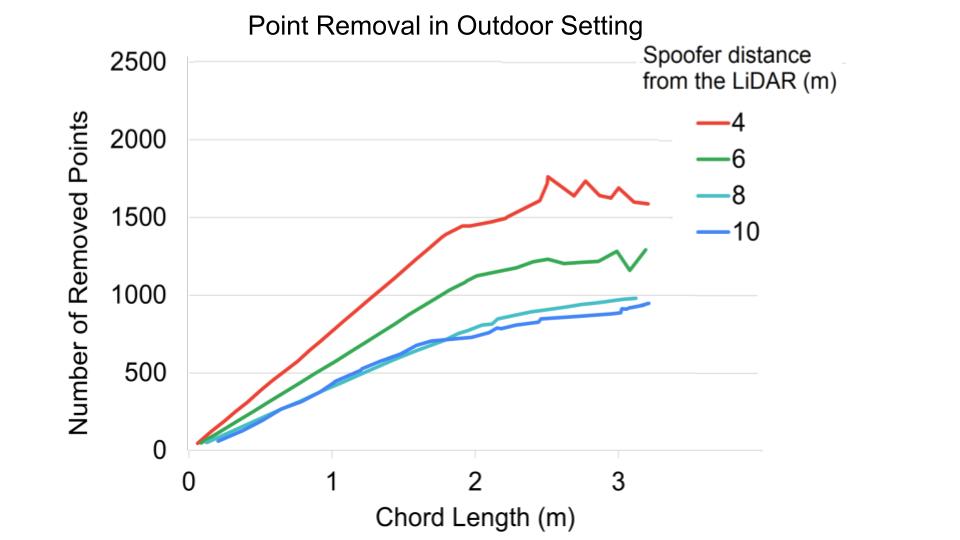}
   \end{minipage}\hfill
   \caption{Number of removed 3D cloud points in the attack region at increasing attack angles for indoor (top) and outdoor scenario (bottom) with a Velodyne VLP-16 LiDAR sensor. 
   }
    \label{fig:attack_capability_indoor}
    \vspace{-0.1in}
\end{figure}

We first characterize the maximum attack angle in an indoor environment, where the LiDAR is 2.5m in front of the spoofer. The evaluation of the attacker capability for farther distances (up to 10 m) is detailed in \S\ref{subsec:eval} in the outdoor scenario. The LiDAR can capture the maximum number of points due to the controlled environment and also facilitates measurement of the attacker's ability to remove points.
By adjusting the spoofer firing timing, we aimed to find the attack angle range at which the attack successfully and stably removes cloud points.
Starting from 0$^{\circ}$ attack angle,
we increase the number of spoofing points (100 for each step) and measure the total number of resulting removed points until the number of removed points stops increasing. We recorded the total number of removed points in the attack angle and show the results in Figure~\ref{fig:attack_capability_indoor} (top). We note that, when the attack angle is less than 45$^{\circ}$, the number of removed points increases linearly, removing all the points in the region. More specifically, since the horizontal resolution for VLP-16 LiDAR is $0.2^{\circ}$, we can calculate the ratio of the number of removed points over the attack angle as $16/0.2=80$ (i.e. 80 removed points per degree). We conclude that the attacker is able to stably remove all the points in an attack region of around 45$^{\circ}$ and the removal attack gets less stable for further angles. We thus consider 45$^{\circ}$ the maximum attacker capability in controlled scenarios. Noticed that, by selecting a subset of the attack angle (e.g., 10$^{\circ}$), the attacker is able to place the spoofer aside compared to the position of the target object to remove (e.g., $45-10=35^\circ$).

\parhead{Physical Constraints.} The removal attack capability is limited by two factors: 1) the limited receptive field of the photodiodes of the LiDAR system and 2) the limited output power of the laser diodes in our single-spoofer setup. The constrained receptive field of the photodiodes is due to the optical components inside the LiDAR only receiving reflections from certain directions. Therefore, when the spoofer is out of the receptive range, the spoofed signals are not received by the LiDAR system, so the removal attack fails. The limited output power of the laser diodes generate a diverged circular laser spot, even with the optical support to converge it. The intensity of the spoofed pulses are lower when they get nearer to the edge of the spot. When the signal is too weak to spoof the points, the removal attack fails. Also, due to the decaying intensity as the distance between the spoofer and the LiDAR increases, the removal attack capability also decreases. Many of these limitations are therefore caused by the spoofer design.

\subsection{Removing Selected Obstacles}
\label{subsec:removing_cone}
The goal of this evaluation is to analyze the attack's ability to completely remove a target object from LiDAR's perception at different distances. To achieve this, we performed a reduced-scale real-world experiment in a controlled environment. We consider a standard traffic cone as the target obstacle. We then observe the number of points the attack successfully removes at increasing attack magnitudes (i.e., number of spoofed points).

\begin{figure}[t]
  \centering
    \includegraphics[width=0.45\textwidth]{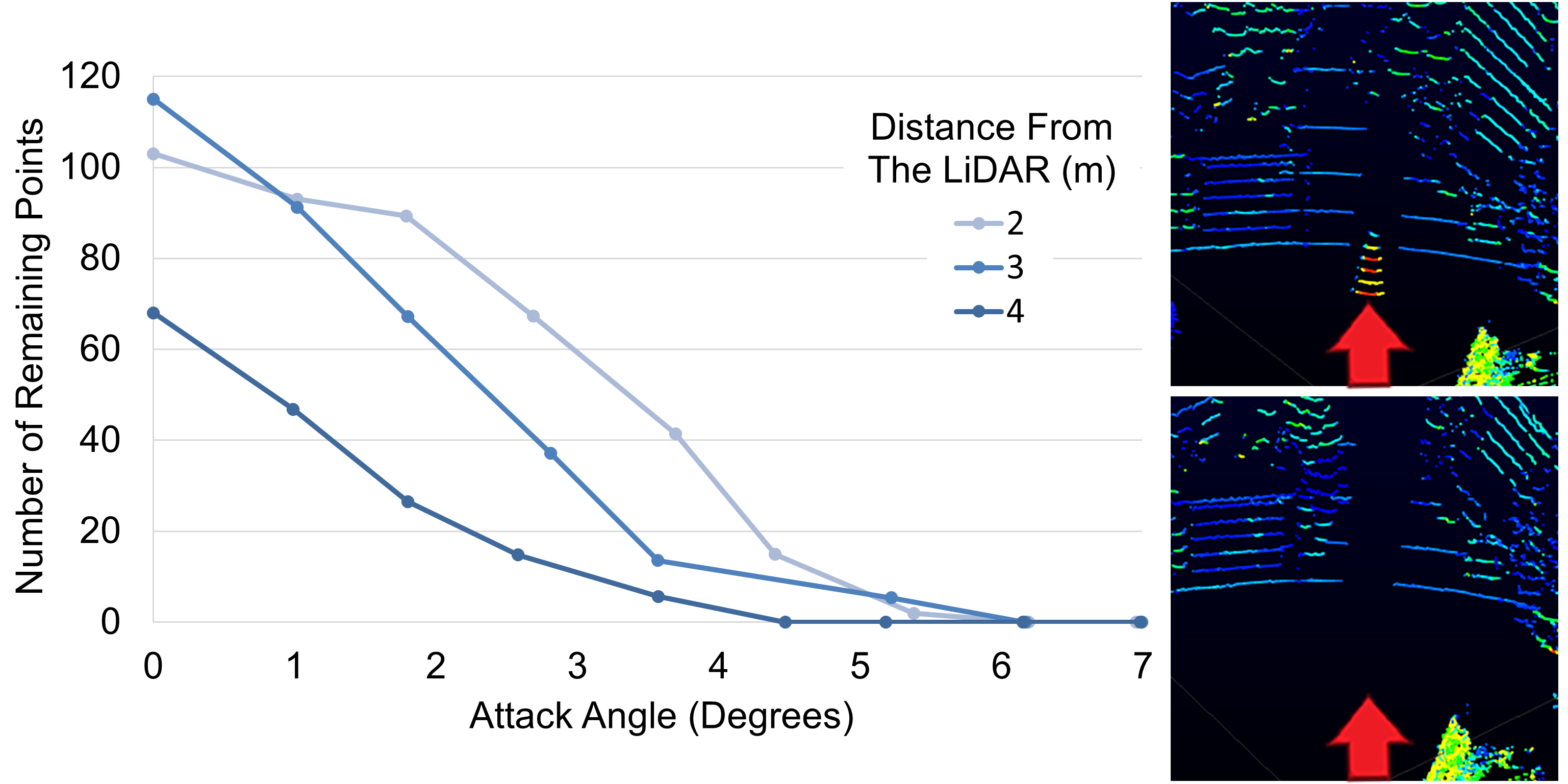}
    \caption{Removed cloud points of a target traffic cone at increasing attack angles (indoor scenario). The traffic cone is located 2, 3 and 4 meters away from the victim LiDAR.}

    \label{fig:traffic_cone}
\end{figure}

\parhead{Experimental Setup.}
In this experiment, we use the same attack setup described in~\S\ref{subsec:attacker_capability}. The traffic cone is located at 2 m, 3 m, and 4 m away from the LiDAR, while the spoofer is located on beside the target object as depicted in Figure~\ref{fig:traffic_cone}.
As explained above, every increment in the attack angle corresponds to an increment of the number of spoofed points by 100 using our spoofer device. Every attack angle is maintained for an average of 20 LiDAR rotations (or \textit{frames}), then we increase the attack angle until all the cloud points of the cone are removed. In this setting, the cone consists of an average of 120, 100, and 70 cloud points at distances 2 m, 3 m, and 4 m, respectively. We then calculate the number of points removed from the actual cone cloud points at different attack magnitudes and the spoofing attack angles.

\parhead{Evaluation Results.}
As shown in Figure~\ref{fig:traffic_cone}, a $4^{\circ}$ attack angle was able to remove an average of 85 points, constituting more than 80\% of the points at every distance considered in the experiment. As expected, the attack angle required to remove the traffic cone from LiDAR perception decreases with increasing distance as the number of points to be spoofed also decreases. The attack was able to completely remove the traffic cone point cloud with an attack angle of 6$^{\circ}$ in the case of 2 m and 3 m distances and an angle of 4$^{\circ}$ in the case of 4 m distance.

\subsection{Effect on AVs Perception Models}
\label{subsec:kitti_synthesis}

The perception module in AVs is the core module that leverages machine learning models such as Convolutional Neural Networks (CNNs) for object detection. The output of this module directly influences the AVs safety-critical driving decisions, such as collision avoidance.

We therefore explore how PRA can affect obstacles confidence score predicted by various LiDAR-based perception modules at increasing attack angles. To achieve this, we synthesized our attack and simulated the on-road scenario with real-world LiDAR traces.

\parhead{Experiment Setup.} We select 200 target obstacles (100 pedestrians and 100 vehicles) at increasing distances (from 6 m to 28 m) and orientations from the LiDAR in the KITTI dataset~\cite{kitti_3d}. KITTI is a popular dataset for benchmarking AD research,  commonly used for evaluating the performance of AVs perception modules. We randomly select target obstacles with high confidence score (0.95 average confidence score for vehicle object and 0.71 for pedestrian object on Apollo's CNN detection model) compared to 0.83/0.64 for randomly chosen vehicle/pedestrian objects. We use these samples in all the synthesized analysis of this work to emulate the worst attacker condition when trying to remove a high-confidence obstacle from the scene.

\begin{figure}[t]
  \centering
    \includegraphics[width=0.45\textwidth]{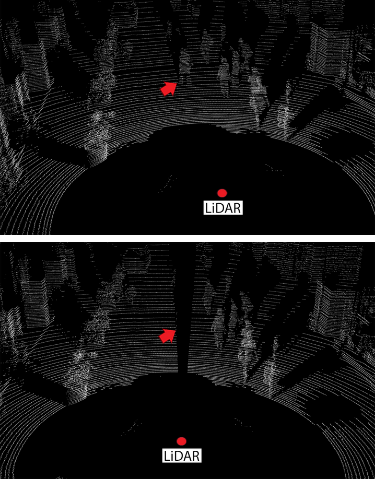}
    \caption{Example of PRA synthesis on a KITTI dataset sample. (Top) The ground truth data shows a pedestrian in front of the vehicle (the red spot indicates the LiDAR location). (Bottom) The pedestrian point cloud removed by the attack in the synthesized data (4$^{\circ}$ attack angle).}
    \label{fig:PRAsynthesis}
\end{figure}

To synthesize the point cloud, we first locate the attack region based on the position of the target pedestrian/vehicle. Then we increment the attack angle by 1 degree each step following the methodology defined in~\ref{subsec:removing_cone} until the obstacle is fully covered, as shown in Figure~\ref{fig:PRAsynthesis}. We evaluate the resulting attack traces with Apollo 5.0~\cite{apollo}, PointPillars~\cite{lang2019pointpillars}, and Autoware~\cite{autoware}.

For Apollo 5.0 and PointPillars, we collect the confidence score values. For Autoware, we extract the clustering results.

\begin{figure}[t!]
   \begin{minipage}{0.238\textwidth}
     \centering
     \includegraphics[scale=0.18]{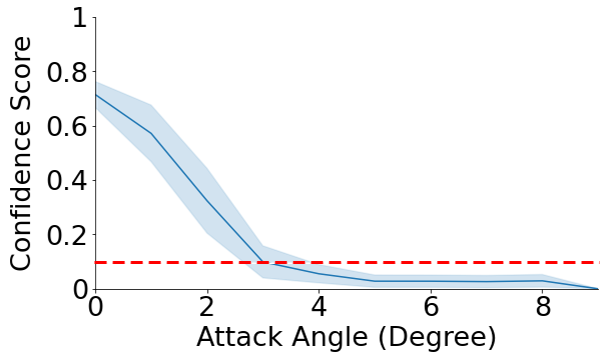}

     \text{(a)}

   \end{minipage}\hfill
    \begin{minipage}{0.238\textwidth}
     \centering
     \includegraphics[scale=0.18]{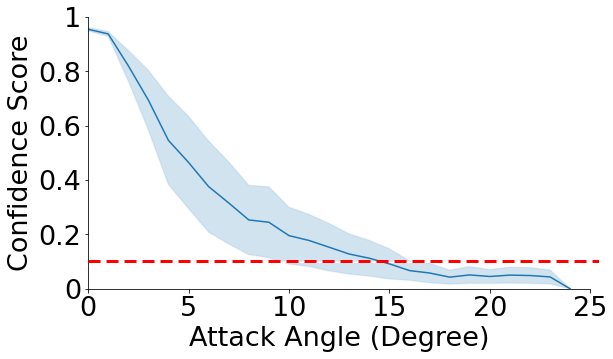}

     \text{(b)}

   \end{minipage}\hfill  
   \begin{minipage}{0.238\textwidth}
     \centering
     \includegraphics[scale=0.18]{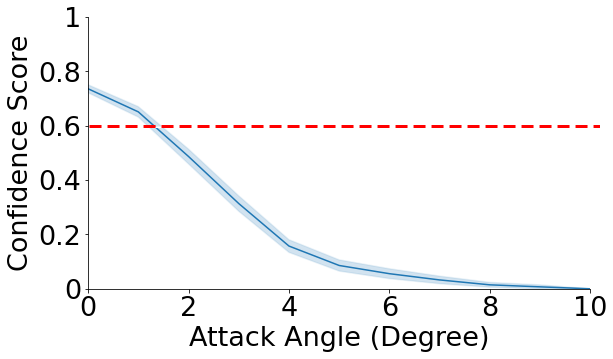}

    \text{(c)}

   \end{minipage}\hfill
   \begin{minipage}{0.238\textwidth}
     \centering
     \includegraphics[scale=0.18]{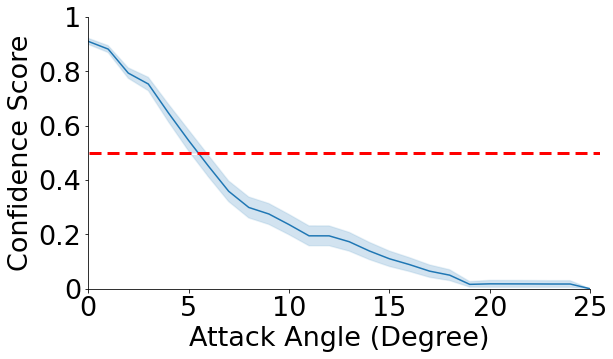}

    \text{(d)}
   \end{minipage}\hfill
   \begin{minipage}{0.238\textwidth}
     \centering
     \includegraphics[scale=0.18]{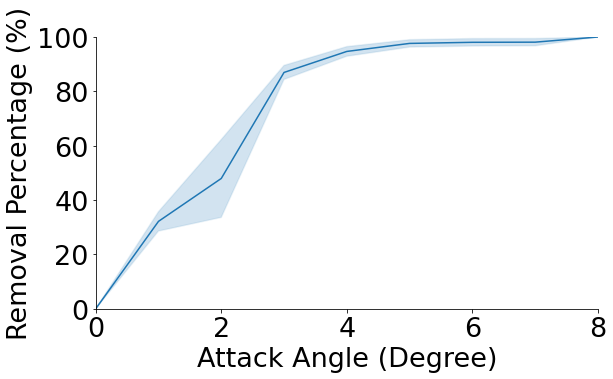}

     \text{(e)}

   \end{minipage}\hfill
   \begin{minipage}{0.238\textwidth}
     \centering
     \includegraphics[scale=0.18]{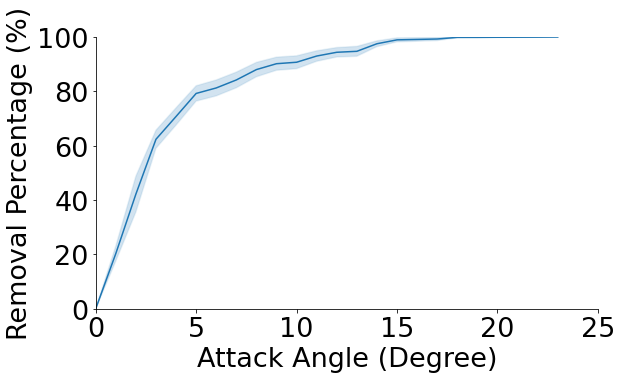}

     \text{(f)}
   \end{minipage}\hfill
      \begin{minipage}{0.238\textwidth}
     \centering
     \includegraphics[scale=0.18]{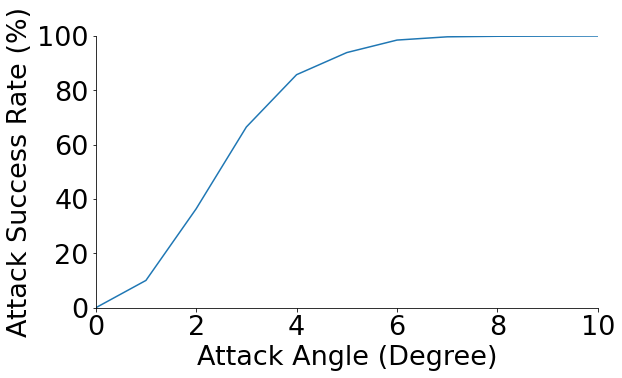}

    \text{(g)}
   \end{minipage}\hfill
   \begin{minipage}{0.238\textwidth}
     \centering
     \includegraphics[scale=0.18]{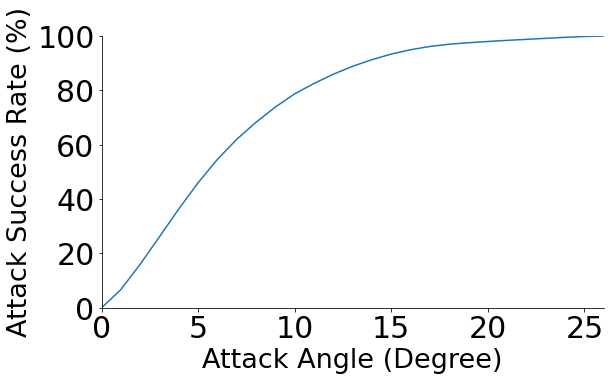}
     \text{(h)}
   \end{minipage}\hfill
        
    \caption{(a) and (b) show the confidence score of Baidu Apollo on pedestrians and vehicles at different angles and (c), (d) show the scores of PointPillars. (e) and (f) represent the percentage of removed cluster points for Autoware in the case of pedestrian and vehicles, respectively. (g) and (h) show the the attack success rate w.r.t angle in the case of pedestrian and vehicle objects respectively (100 samples for each class).}
     \label{fig:apollo_kitti_pedestrian}
\end{figure}

\parhead{Evaluation Metrics.}
To evaluate our experimental results, we considered two different metrics: confidence score and number of removed points.
Some object detection models provide thresholds for confidence scores. This allows for discarding low confidence objects (potential false positives). We consider the default thresholds used by two of our tested models (0.1 in Apollo~\cite{apollo} and 0.5 and 0.6 for vehicle and pedestrian detection in PointPillars~\cite{lang2019pointpillars} respectively). We then consider an attack successful for those two models when the model confidence score for the pedestrian (or vehicle) falls below the default threshold. 
For Autoware instead, we consider the results of clustering in terms of the number of removed points in the detected cluster. In this case, we consider the attack successful when all the points are removed from the cluster.

Finally, we define additional metrics that leverage the number of removed points variation over attack angle for obstacles to evaluate the overall attack success rate.
We calculate this value as the ratio of the average number of removed points to the average number of cloud points of the target objects in the KITTI dataset.
We consider an attack successful for all three models when the average of the fraction of removed points reaches zero (i.e., all the target obstacle points have been removed).

\noindent\textbf{Results.} 
Figure~\ref{fig:apollo_kitti_pedestrian} shows that the attack drops the confidence score below the required threshold (red dashed line) at an attack angle of 15$^{\circ}$ and 6$^{\circ}$ for Apollo and PointPillars respectively in the case of a vehicle obstacle and 3$^{\circ}$, 2$^{\circ}$ in the case of a pedestrian obstacle. In the case of Autoware, the entire cluster was removed at a 23$^{\circ}$ attack angle for a vehicle obstacle and 8$^{\circ}$ for a pedestrian obstacle. 
Finally, we show the required attack angle to fully remove the obstacles in (g) and (h) based on the KITTI dataset ground truth (8$^{\circ}$ for pedestrian obstacles and 24$^{\circ}$ for vehicle obstacles).
We also noticed when reporting the confidence scores for Apollo, the remaining spurious point cloud was sometimes clustered as a part of nearby obstacles or the point cloud was clustered as multiple obstacles. In such cases, we report the confidence score as zero since it failed to detect the obstacle. The results demonstrate that the perception models confidence drop quickly as the attack angle increases and the models fails to detect an obstacle even when the obstacle point cloud is not fully removed, especially for pedestrians.

\subsection{Impact on Fusion}
\label{subsec:fusion}
This section demonstrates that our removal attack is robust against state-of-the-art camera-LiDAR fusion models for object detection, object localization, and tracking~\cite{autoware,apollo}. 
Fusion in AVs helps compensate for the accuracy limitations of using individual sensors, and it can provide additional robustness against naive black-box attacks on AV perception modules~\cite{HallyburtonFusionSecurity}. 
We consider the same experimental setup as in~\S\ref{subsec:kitti_synthesis} using our 200 high-confidence samples to evaluate our attack against three popular camera-LiDAR fusion architectures: 

\begin{itemize}
\item Frustum-ConvNet (FC)~\cite{fusionFrustum}. This Cascaded Semantic Level Fusion model takes the region proposals from the camera image and creates frustum level feature vectors from the LiDAR points for each region proposal. The model spatially fuses these frustum features to estimate an oriented 3D bounding box.

\item AVOD \cite{fusionAVOD}. This Feature Level Fusion extracts individual feature maps from the raw camera and LiDAR data and combines them.

\item Autoware Integrated-Semantic Level Fusion~\cite{autoware}. The model fuses the feature outputs from independent detection stacks of camera and LiDAR sensors by back-projecting the detected object from the LiDAR onto the image space. Then it evaluates if there is at least a 50\% overlap between the two detections to confirm an obstacle.
\end{itemize}

\parhead{Evaluation Metrics.} We use the detection rate as a metric to evaluate PRA on the three fusion models. 

We consider two analyses. 
In the first analysis, the Intersection-over-Union (IoU) evaluation is performed on the default (DEF) thresholds for each model (0.7 for cars and 0.5 for pedestrians in the case of AVOD and Frustum-ConvNet, and a 50\% overlap in Autoware Integrated-Semantic Level Fusion). In the second analysis (AVE), the evaluation is performed for all the possible IoU threshold values over 3D bounding box predictions for each fusion model (0.1 - 0.9 for FC and AVOD, 10\% to 90\% for Autoware). Similar to~\S\ref{subsec:kitti_synthesis}, we increment the attack angle by 1 degree until the obstacle is completely removed, which corresponds to 8$^{\circ}$ for pedestrian obstacles and 24$^{\circ}$ for vehicle obstacles according to Figure~\ref{fig:apollo_kitti_pedestrian}.

\begin{table}[h!]
 \caption{Object detection rates on fusion models at increasing attack angles for pedestrian and vehicle target obstacles\footreference{website}.} \label{tbl:objectFusion}
\centering
{\small
\scalebox{0.77}{
 \begin{tabular}{@{}c|cc|cc|cc@{}}
 \toprule
 \multirow{2}*{\begin{tabular}{c}\textbf{\footnotesize{Attack}}\\
                                 \textbf{\footnotesize{Angle (\textdegree)}}\\
                                 \end{tabular}}&
  \multicolumn{2}{c|}{\textbf{\footnotesize{FC}}}  & \multicolumn{2}{c|}{\textbf{\footnotesize{AVOD}}}  & \multicolumn{2}{c}{\textbf{\footnotesize{Autoware}}}\\  
  \cmidrule(l|{.2em}r|{.2em}){2-3} \cmidrule(l|{.2em}r|{.2em}){4-5} \cmidrule(l|{.2em}r|{.2em}){6-7} 
  & \textbf{\footnotesize{DEF (\%)}}&\textbf{\footnotesize{AVE (\%)}}&\textbf{\footnotesize{DEF (\%)}}&\textbf{\footnotesize{AVE (\%)}}&\textbf{\footnotesize{DEF (\%)}}&\textbf{\footnotesize{AVE (\%)}}\\ 
   \midrule
  \multicolumn{7}{c}{\textbf{Pedestrian Detection Rates}}\\
\midrule
0 & 73 & 58.3 & 80 & 64.6 & 63 & 49.3 \\
 1 & 50 & 48.7 & 66 & 50.5 & 56 & 43.7 \\
 2 & 34 & 37.2 & 36 & 27.2 & 41 & 35.6 \\
 3 & 17 & 24.0 & 9  & 8.3  & 12 & 13.6 \\
 4 & 4  & 15.4 & 4  & 3.2  & 4  & 4.8\\
 5 & 6  & 12.7 & 3  & 2.4  & 2  & 1.8 \\
 6 & 2  & 11.0 & 3  & 2.4  & 0  & 0.0 \\
 7 & 2  & 10.8 & 3  & 2.4  & 0  & 0.0 \\
 8 & 4  & 11.1 & 3  & 2.4  & 0  & 0.0 \\
 \midrule
   \multicolumn{7}{c}{\textbf{Vehicle Detection Rates}}\\
  \midrule
0 & 72 & 78.2 & 82 & 81.1 & 44 & 42.8 \\
 3 & 56 & 69.1 & 72 & 72.8 & 23 & 25.8 \\
 6 & 37 & 53.8 & 43 & 44.2 & 15 & 16.3 \\
 9 & 28 & 43.9 & 21 & 26.1 & 6  & 6.9  \\
 12& 15 & 35.0 & 6  & 15.4 & 3  & 4.7  \\
 15& 8  & 29.7 & 6  & 12.1 & 1  & 0.8  \\
 18& 7  & 29.6 & 8  & 12.0 & 0  & 0.3  \\
 21& 7  & 28.7 & 7  & 11.6 & 0  & 0.1  \\
 24& 6  & 29.0 & 8  & 12.1 & 0  & 0.0    \\
\midrule
\multicolumn{7}{l}{\footnotesize{* 0$^{\circ}$ attack angle corresponds to a scenario with no attack.   }}\\
\end{tabular}
}
}
\end{table}

\parhead{Results and Observations.}
Table~\ref{tbl:objectFusion} shows the fusion results for pedestrian and vehicle target obstacles. 
 In the case of pedestrian obstacles, we observe that the detection rate in DEF drops by 69\% for FC, and 76\% for AVOD at an attack angle of 4$^{\circ}$ (half of the required attack angle to remove the obstacle). The AVE analysis, on the other hand, shows that the detection rates on average drop by 43\% for FC, and 61\% in AVOD at a 4$^{\circ}$ attack angle.
At a 15$^{\circ}$ attack angle, the detection rate of vehicle obstacles drops by 59\% in FC and 76\% in AVOD for the default IoU thresholds. The AVE analysis instead shows an average drop of 48.6\% and 69\% in detection rate for FC and AVOD, respectively. 

We observe that the detection rates in Autoware fusion drop to 0\% at angles immediately before full obstacle removal in both DEF and AVE analysis. This is because the LiDAR detected obstacle does not overlap with the camera detected obstacle by at least 50\%.

Our results demonstrate that Autoware's fusion always fails to fuse detected objects regardless of the IoU threshold when almost all the cloud points are removed.

In addition, our results show that with an attack angle of 8$^{\circ}$ and 24$^{\circ}$ for pedestrian and vehicle obstacles, respectively, the detection rate of our high-confidence objects drops by at least 43\% in all three fusion models. This confirms that our attack can cause severe performance drops and failures even if fusion is implemented.

%% file: eval.tex
\section{Physical Removal Attack Evaluation}
\label{subsec:eval}
To conduct the removal attack in the real world, we consider the following scenario where the autonomous vehicle is driving forwards, approaching a static (or a moving) obstacle in the center of the AV trajectory, with the attacker's spoofer located on the side of the road. The goal of the attacker is to remove the obstacle from the 3D map of the LiDAR sensor. Since the LiDAR scans are not uniformly distributed along the longitudinal axis and become sparser at increasing distance, we propose to use chord length to replace attack angle as the evaluation metric in this particular setting (Figure~\ref{fig:optical_diagram}). The chord length defined in~\ref{subsec:overview}, which depends on the target obstacle's dimensions, the distance between the obstacle and the LiDAR, and attack angle, better captures the amount of cloud points removed in the attack region. For example, if the victim LiDAR is far away from the obstacle, the attacker would need to spoof fewer points in the cord length (namely a small $\Delta \theta$) compared to the scenario with the LiDAR in proximity of target obstacle to remove.  

In this section, we consider two sets of evaluation: (1) the attacker capability in outdoor scenarios and (2) the impact of our attack on the driving decisions by evaluating end-to-end scenarios on prediction, control, and planning using the industry-grade simulator LGSVL. The moving AV scenarios will be shown in~\S\ref{subsec:tracking}.

\subsection{Outdoor Scenario}
\label{subsec:outdoor_exp}

\begin{figure}[t!]
  \centering
    \includegraphics[width=0.48\textwidth]{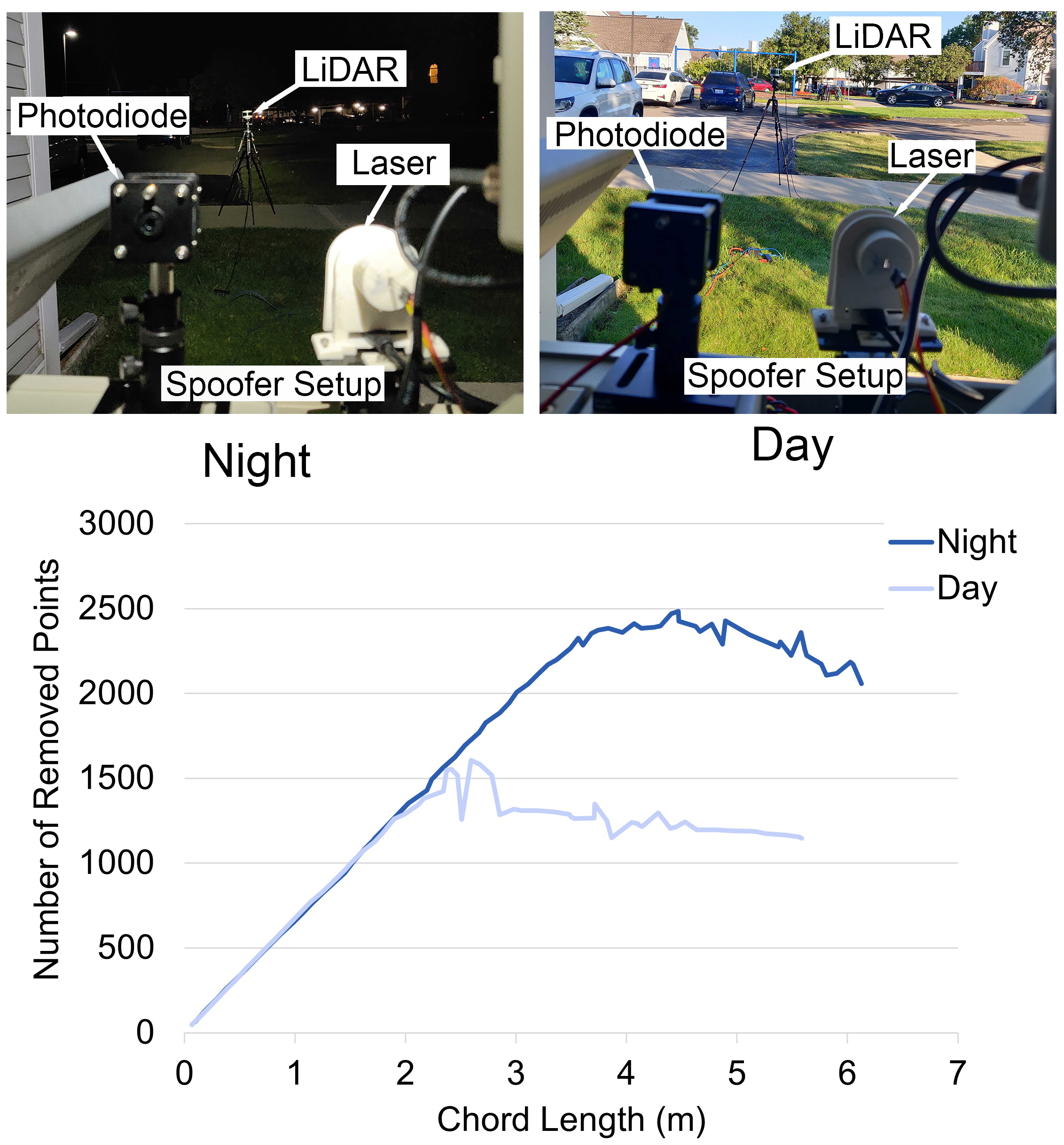}

    \caption{Number of removed cloud points comparison during the day (top right) and night (top left). The distance between the LiDAR and spoofer is 5 m. }

    \label{fig:attack_capability_day_night}
\end{figure}

\parhead{Attacker Capability at Increasing Distances.}

Since we noticed that the attacker capability is symmetrical on two sides of the spoofer during the attacker capability experiments conducted in~\S\ref{subsec:attacker_capability}, for the experiments evaluating the attack robustness outdoor we only conduct the removal attack on one side of the spoofer. 
We then measured the number of removed points by conducting the removal attack with the LiDAR placed at different distances from the spoofer during the daytime (up to 10 m). The results in Figure~\ref{fig:attack_capability_indoor} (bottom) show that, as the distance from the LiDAR increases, the attack capability of removing points decreases. This is mainly due to the limitations of the spoofer setup, as described in~\S\ref{subsec:attacker_capability}. For instance, the intensity of the spoofed signals decreases as the distance increases, weakening the overall attacker capability.
\parhead{Different Light Conditions.}
For different light conditions, we measured the number of removed points by conducting the removal attack, with the spoofer 5 m away from the LiDAR during day and night. The evaluation results in Figure~\ref{fig:attack_capability_day_night} show that the attacker capability is reduced under daylight condition. One potential reason observed in the literature is that natural light also contains IR components at the same wavelength as the LiDAR signals (905 nm) that may interfere with sensor functioning~\cite{lambert2020performance}.

\begin{figure}[t]
  \centering
    \includegraphics[width=0.48\textwidth]{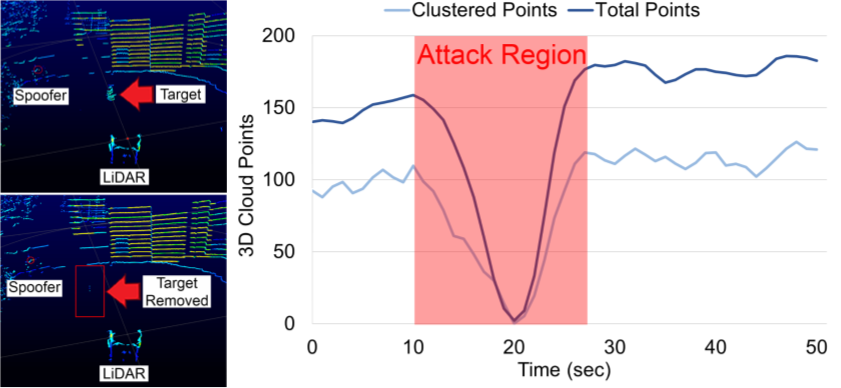}
    \caption{(Left) The 3D point cloud (top) with and (bottom) without the attack. (Right) The decimation of the genuine cloud points during the attack. When the pedestrian walks in the attack region (10 sec) and leaves (27 sec) (red area) the cloud points of the pedestrian and related cluster in Autoware are reduced to zero\footreference{website}. }

    \label{fig:ped_confidence}
\end{figure}

\noindent\textbf{Removing a Moving Obstacle.}
In this experiment, we evaluate the removal attack in the following scenario: a pedestrian walking across in front of a stopped autonomous vehicle.
With the spoofer deployed 8 m away from the LiDAR on the side of the road aiming to remove the walking pedestrian, who is 4 m in front of the LiDAR. The LiDAR is placed on top of the victim vehicle to simulate the autonomous vehicle setup. We evaluate the captured trace with the Autoware~\cite{autoware} perception module, which is the only publicly available model for processing data from a VLP-16 LiDAR. As shown in Figure~\ref{fig:ped_confidence}, the pedestrian can be fully obscured when walking into the attack region, with its cloud points fully removed. Figure~\ref{fig:ped_confidence} also shows the corresponding detection results from Autoware, in the form of clustered points. We can see that both clustered and total points of the pedestrian  are reduced to zero.

%% file: simulation.tex
\subsection{Impact on Driving Decisions} 
\label{sec:sim}

\begin{figure*}[t!]
   \begin{minipage}{0.32\textwidth}
     \centering
     \includegraphics[width=5.7cm, height=2.8cm]{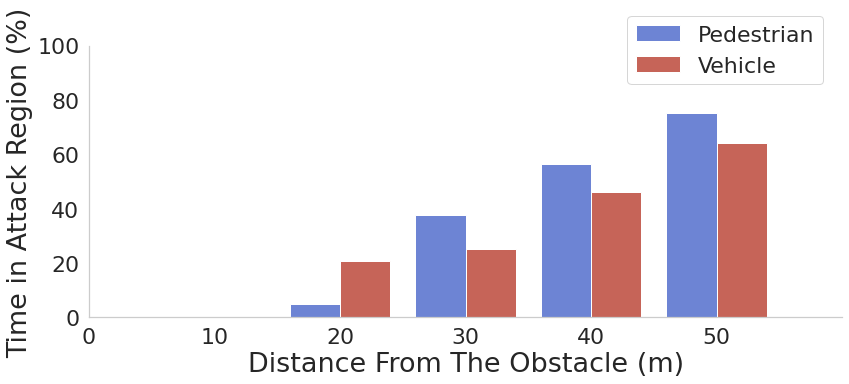}
    \text{(a)}
   \end{minipage}\hfill
   \begin{minipage}{0.275\textwidth}
     \centering
     \includegraphics[width=5cm, height=2.8cm]{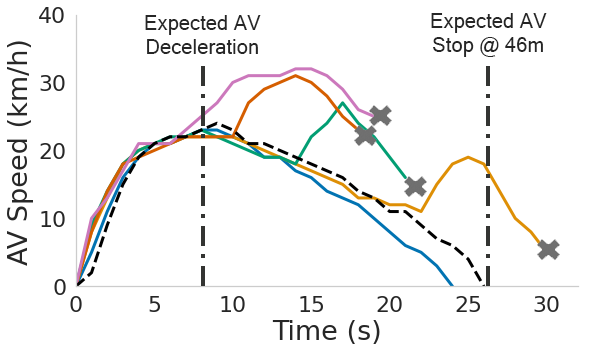}
     \text{(b)}
   \end{minipage}\hfill
   \begin{minipage}{0.405\textwidth}
     \centering
     \includegraphics[width=7.0cm, height=2.8cm]{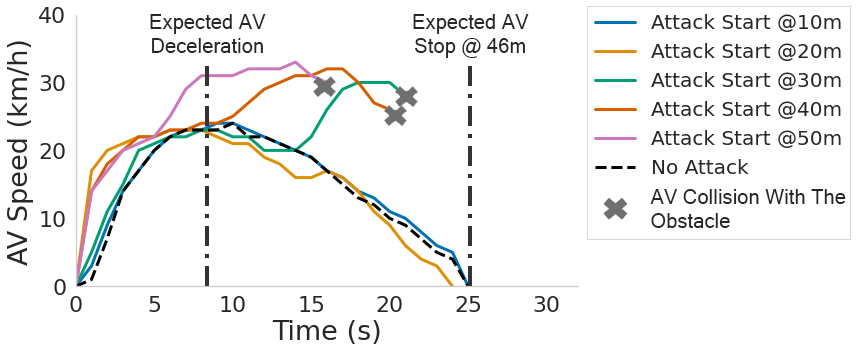}
     \text{(c)}
   \end{minipage}\hfill
    \begin{minipage}{0.32\textwidth}
     \centering
     \includegraphics[width=5.7cm, height=2.8cm]{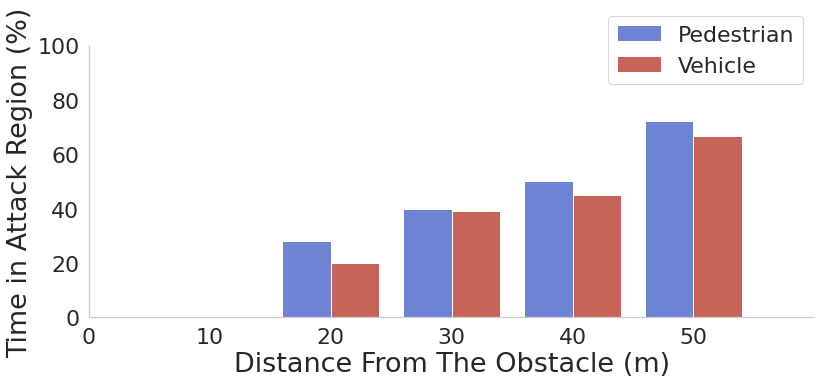}
     \text{(d)}
   \end{minipage}\hfill  
   \begin{minipage}{0.275\textwidth}
     \centering
     \includegraphics[width=5cm, height=2.8cm]{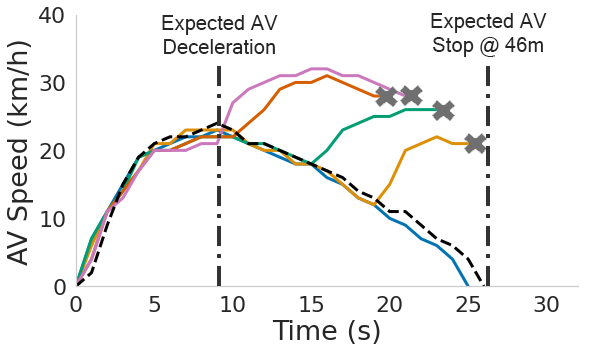}
    \text{(e)}
   \end{minipage}\hfill
   \begin{minipage}{0.405\textwidth}
     \centering
     \includegraphics[width=7.0cm, height=2.8cm]{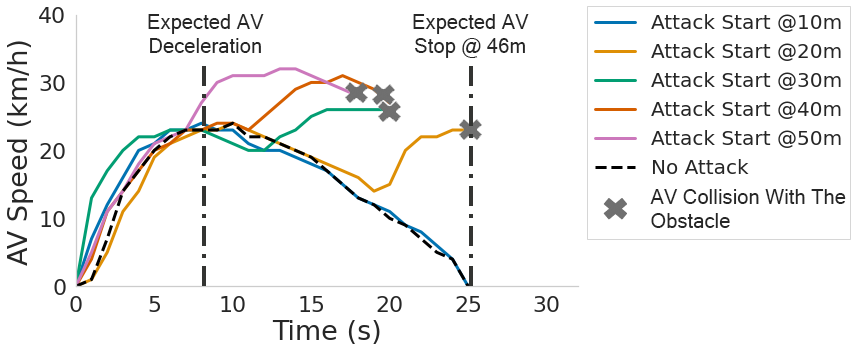}
    \text{(f)}
   \end{minipage}\hfill
    \vspace{-0.1in}
    \caption{LGSVL simulation results for 5$^{\circ}$ (top) and 10$^{\circ}$ (bottom) removal attack angles. (a) and (d) shows the amount of time (\%) the target obstacle remains in the attack region over the AV's entire route time. (b), (c), (e), and (f) show the AV speed change at different attack start distances.}
     \label{fig:simulationTrejectory}
     \vspace{-0.1in}
\end{figure*}
To illustrate the consequences of the attack in AD settings, we simulate our PRA in an AD simulator. The simulation goal consists of evaluating changes in the AV trajectory and speed under different scenarios where the adversary launches our removal attack with the intent to hide an obstacle in the victim's route.
We perform our end-to-end attack evaluation on Baidu Apollo using the LGSVL simulator~\cite{rong2020lgsvl}. LGSVL is an open-source production-grade simulator designed for testing and developing AD frameworks, which supports real-time integration with Apollo and Autoware, and it is widely used in the literature to simulate attack scenarios~\cite{cao2019adversarial, Cao2020invisible, shen2020drift}. The LGSVL simulator also uses semantic fusion on objects tracked individually from the camera and LiDAR as part of its object detection stack.

To simulate PRA, we synthesize the sensor input changes corresponding to our attack on the input of the perception module in Apollo and direct the resulting output to the simulator to evaluate the AV control decisions in real-time. Specifically, we model the LiDAR point cloud in the simulated HDL 64 LiDAR traces to match our attack traces on the VLP-16 as in previous works~\cite{cao2019adversarial}. We also assume an ideal scenario where all cloud points in the attack region are removed.

To extend the realism of the simulated attack, we start the attack when the AV is at different distances away from the spoofer and target obstacle to simulate the attack capabilities measured in \S\ref{subsec:outdoor_exp}. By such means, we can measure how different attack capabilities affect driving decisions.

\parhead{Simulation Settings.} We explore autonomous vehicle behavior under different scenarios during our removal attack. For each scenario, we consider the victim's AV moving at constant acceleration from 0 to 32 km/h on a single-lane road, which is the maximum speed limit from Apollo's planning configuration for a single-lane map in LGSVL. The AV approaches two types of static obstacles (a car and a pedestrian) located in different positions along a crosswalk (e.g., along the AV's trajectory). The selected scenarios simulate a pedestrian crossing the road, or a car at rest, such as at a traffic light or in the proximity of an intersection. To further reflect the distance constraints demonstrated in the physical experiments, we start simulating the attack when the AV's distance to the obstacle is 10, 20, 30, 40 and 50 meters, respectively. We use Apollo version 5.0 and configure the LGSVL simulated LiDAR with 64 lines, 10 FPS, and 0.1 degree angle resolution to match the capability of Velodyne HDL 64 LiDAR. 
Our evaluation considers each obstacle located on the crosswalk at one of 5 distinct positions at 1 m apart, covering the entire road width. We consider the spoofer located on the right side of the road, in proximity to the crosswalk, and we consider 5- and 10-degree attack angles for each scene. 

From the simulations, we then extract the AV speed and trajectory changes over time depending on the time the obstacle remains in the attack region.

\begin{figure}[t!]
    \centering
    \includegraphics[width=\linewidth]{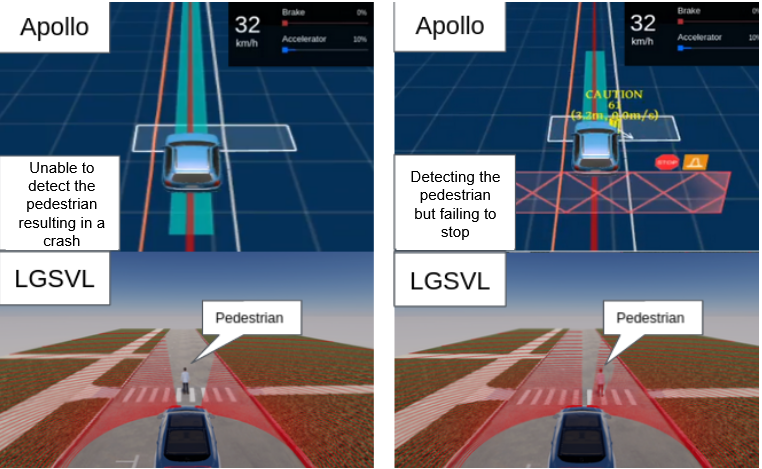}
    \caption{PRA simulated on LGSVL. (Left) With 10$^{\circ}$ attack angle, the vehicle is unable to detect the pedestrian. (Right) For a small attack angle (5$^{\circ}$), the vehicle detects the pedestrian too late to timely stop.
    }
    \label{fig:sim_overview}
\end{figure}

\parhead{Simulation Results.}
Figure~\ref{fig:simulationTrejectory} demonstrates that our PRA 
can lead to severe consequences and endanger the victim AV (e.g., by colliding with obstacles on the road). Figure~\ref{fig:simulationTrejectory} (a) and (d) show that by starting the attack at different distances, the attacker can remove the target obstacles for different time periods (based on the size of the obstacles and the attack angle).
Figures~\ref{fig:simulationTrejectory} (b), (c), (e), and (f) show that though the obstacle is only removed for a limited amount of time, it will cause the AV to accelerate and collide with the obstacle. This happens because without the attack, the victim AV is expected to accelerate to reach the preset AV speed (32 km/h) at 46 meters, then uniformly decelerate and stop before reaching the obstacle. Therefore, when the attack starts and the target obstacle is removed, the victim AV accelerates to reach the preset speed instead of decelerating.
Though the obstacle might be perceived again, such as in the case of smaller attack angles, we observe that the AV can still collide with it unless the distance between the AV and the obstacle $d$ is such that $d < \textit{v}^{2}/(2 \cdot a)$, where \textit{v} is the velocity of the AV when the obstacle reappears and \textit{a} is the AV's deceleration rate.
Figure~\ref{fig:sim_overview} shows the consequent object detection of Apollo with the simulated attack in LGSVL.

Our results show that PRA can cause AVs to fail to brake and stop before colliding with an obstacle even if the obstacle is inside the attack region for only 40\% of the entire AV's route time. Therefore, we can conclude that, even with limited capabilities of attack angles, our removal attack can lead to severe consequences for AVs like collisions.

\parhead{Limitations of the Simulation Scenario.}
In our evaluation we notice that, without the attack, the vehicle stops at more than 10 meters in front of the obstacles due to the conservative Apollo planning algorithms. Thus, our removal attack simulation does not start when the car is 10 meters away from the obstacle, limiting our evaluation for scenarios where the distance between the vehicle and the obstacle is greater than 10 meters.
Another limitation for production-grade AD simulators such as LGSVL~\cite{rong2020lgsvl} and CARLA~\cite{Dosovitskiy17} is that they do not consider fine-grained manipulation of point cloud data for computational efficiency. Due to the requirements of real-time rendering, the LiDAR point cloud renderer usually conducts ray casting for a fixed size of the region (e.g., $20^\circ$ in the horizontal plane for LGSVL). This allows batch computation and speeds up the rendering speed. Due to this limitation, we use the maximum resolution we can reach with a competitive GPU (NVIDIA RTX 3080). Extending the resolution further results in lower rendering FPS, which is not desirable. 

We also observe noise signals around the edge of the region (1-2 degree horizontal angle) that the removal attack targets, which allows a more realistic comparison with the expected noise in the real world. Although the simulator does not model more complex LiDAR sensing such as point cloud interference below 5-degree resolution, our results show that PRA can lead to severe collisions with only a 5-degree attack angle. Though these limitations hinder the analysis for attack impacts given more fine-grained capabilities, the simulation provides evidence of severe consequences of removal attacks.

%% file: tracking_system.tex
\begin{figure}[t!]
    \centering
    \includegraphics[width=\linewidth]{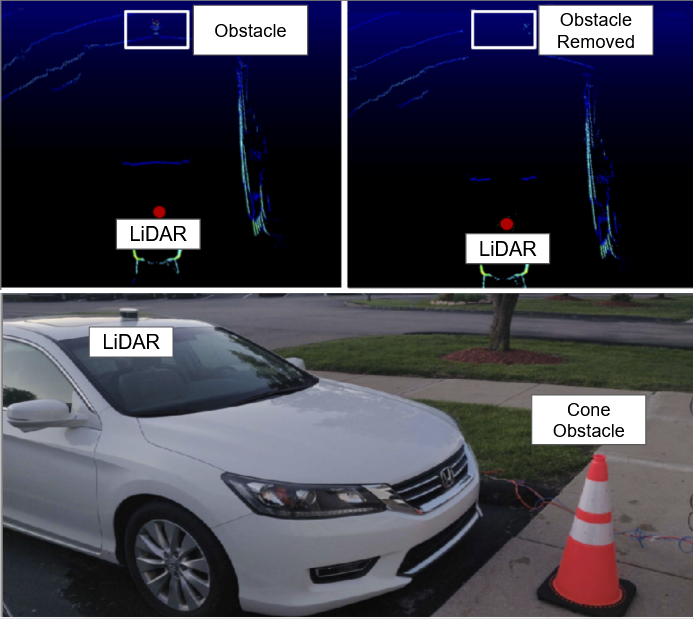}
    \caption{Moving vehicle scenario. (Left-Top) Point cloud without the attack. (Right-Top) Point cloud during the attack. (Bottom) The car setup and the target obstacle (a traffic cone)\footreference{website}.}
    \label{fig:real}
\end{figure}

\begin{table}[t!]\footnotesize
\small
    \centering
   \caption{Success rate of our removal attack in moving scenarios with different threshold levels of removal percentage (RP).}
   \label{tab:moving_target}  
   
    \begin{adjustbox}{max width=\linewidth}
    \begin{tabular}{c|cccc} \toprule
    \multirow{2}{*}{\textbf{Scenario}} & \multicolumn{4}{c}{\textbf{Success Rate (\%)}} \\
         & \textbf{RP(\%)}   & 100 & $\geq$ 95 &  $\geq$ 90   \\\midrule
        Robot &                      & 85.4 & 92.4 & 95.9   \\
        Car &                      & 83.6 & 89.1 & 92.7  \\

          \bottomrule
   \end{tabular}
   \end{adjustbox}

\end{table}

\section{PRA on Moving Vehicles}
\label{subsec:tracking}

In this section, we conduct proof-of-concept evaluation on moving vehicles with the LiDAR placed on top of a robot and a car respectively. In order to conduct the removal of a static obstacle (a traffic cone, and a pedestrian), we build a camera-based tracking system based on the previous work of Cao et al.~\cite{caoautomated}. 
Though attacking moving vehicles introduces additional technical challenges, we demonstrate the feasibility of the attack with a basic tracking system.

\subsection{Technical Challenges and Solutions}
There are two main challenges for attacking a moving vehicle: (1) synchronizing with the LiDAR sensor and, (2) aiming the spoofer at the LiDAR. Therefore, we modify our setup accordingly.

\parhead{Synchronizing with the LiDAR System.} Synchronizing with the LiDAR using a single photodiode at a fixed location is difficult since the LiDAR scan is sparse at distance. This limitation is further aggravated when the AV is moving since the photodiode might fall outside the LiDAR FOV and consequently not capture the laser beams. 

To overcome this challenge we use a larger optical setup (e.g., magnifying lenses) to increase the receptive field so that the LIDAR signals can be captured by the photodiode. However, lens diameter is limited to several tens of centimeters, and is also heavy in weight, reducing our attacker capability over long distances.

\parhead{Aiming with a Camera-based Tracking System.} For aiming the spoofer at the LiDAR, we implement a simple camera-based tracking system with a 3D printer laser holder on top of a robotic turret~\cite{caoautomated}. The system helps aim at the LiDAR but only within the camera's point of view. 
Thus we measure the corresponding position to aim when the LiDAR is at different distances in advance. Then, leveraging the detected bounding box of the tracking model and the true size of the LiDAR, we estimate the distance between the camera and the LiDAR. At last, we look up the stored mapping and aim for the corresponding aiming point for the camera.

\parhead{Tracking System.}
\label{appendix:tracking}
We implement the SSD MobileNet v2 COCO model~\cite{github1} to detect the LiDAR location. 
Although models such as Faster-RCNN ~\cite{ren2015faster} showed better results on our data, SSD MobileNet demonstrated faster speeds required for real-time attacks. This model is based on the principle that layers of a deep network, based on depth, can learn different levels of representations and are built on depth-wise separable filters, which reduce the drastically the required computation.
Howard et al.~\cite{mobilenet1} and Sandler et al.~\cite{mobilenet2} show that a base MobileNet architecture requires significantly less number of parameters to train while showing better than or similar results to models such as AlexNet~\cite{alex_net}, VGG-16~\cite{VGG16}, YOLOv2~\cite{YOLOV2} and other real-time models.
We retrained and tested the model using a custom data, consisting of 720 images of the VLP-16 LiDAR sensor, captured with different settings and scenarios. We used standard data augmentation techniques such as horizontal flipping and random cropping to relatively increase the size of our data set and produce robust models. We were able to achieve 93 mAP (mean average precision) at 50 IoU and a 53 mAP at 75 IoU for the test set at 20 FPS. A better real-time performance requires smoother bounding boxes and tracking through occlusions. To achieve this over continuous frames, we implemented the SORT model~\cite{Wojke2017simple}, which demonstrates the least number of identity switches among online methods, at a rate of approximately 20 Hz (sufficient for real-time). The spatial coordinates of the LiDAR sensor are then extracted from the detected object, with which the calibrations for the attack system's pan-tilt movement are calculated to aim the laser at the victim sensor.

\subsection{Evaluation}
\begin{figure}[t!]
     \centering
     \includegraphics[scale=0.14]{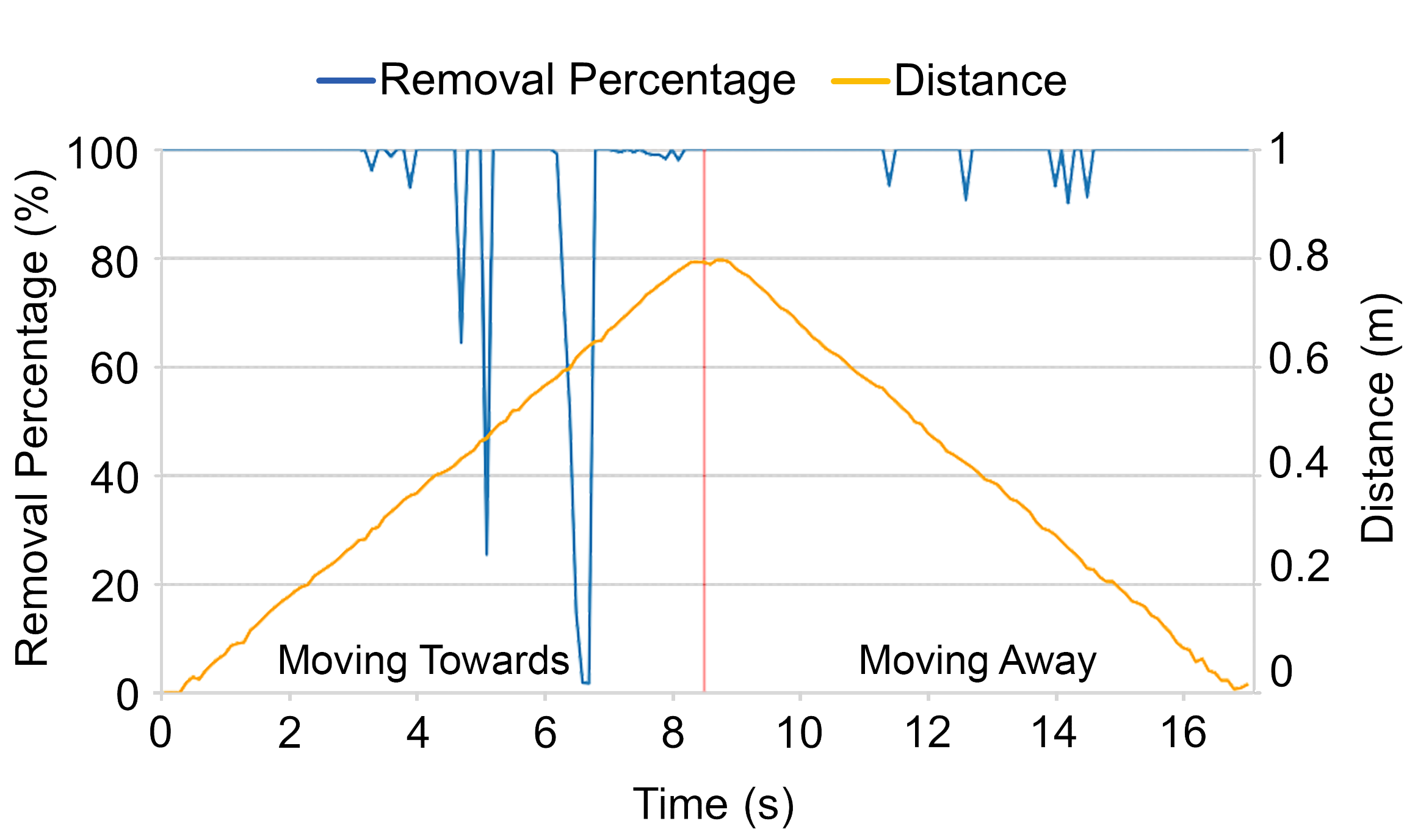}
    \caption{{Cloud point removal percentage of the pedestrian obstacle during our attack. The attack is conducted in two phases: (1) the robot moving towards the target pedestrian obstacle; (2) the robot moving away from the pedestrian.}}
     \label{fig:moving_exp}
\end{figure}

We conduct two experiments where we place the LiDAR on top of a robot and a car. In the robot case, we aim to analyze the performance of the tracking system with a controlled environment, where we program the robot to be moving first towards a pedestrian 5 m ahead, then back to the starting point, at a constant speed. For the car case, we aim to demonstrate the feasibility of attacking the AV in the real world with the proposed tracking system, where we drive the car towards the obstacle from 5 m ahead. Here, due to safety concerns, we use a traffic cone as the obstacle (Figure~\ref{fig:real}). We also place the spoofer 5 m in front of the LiDAR but on the side to simulate a roadside attacker. For the robot, we use a Neato Botvac D85~\cite{Neato}, moving at full speed (0.1 m/s). For the car, we drive at approximately 5 km/h.
\parhead{Results and Observations.}
Our experiments show that, with the tracking system, the attack on a moving vehicle (car and robot) can be feasible in the real world (see Figure~\ref{fig:real}). Figure~\ref{fig:moving_exp} shows the percentage of points removed from the target pedestrian using the robot. We noticed that the obstacle is not fully removed in several frames in both the scenarios, due to the imprecise aiming. The success rate of the attacks given different removal percentage threshold is shown in\\ Table~\ref{tab:moving_target}.

%% file: defense.tex
\section{Defense Analysis}
\label{sec:defense_analysis}
In this section, we first evaluate how current defense strategies against cloud point spoofing and hiding attacks do not mitigate our removal attack, then we propose two defense strategies.

\subsection{Existing Defenses}

\parhead{Spoofing Attack Defense.}
\label{subsec:CARLO}
To detect and mitigate the adversarial sensor attack proposed by Cao et al.~\cite{cao2019adversarial}, Sun et al. proposed CARLO~\cite{sun2020towards}.  CARLO is designed to leverage distance and occlusion information for detecting the injection of fake point cloud patterns. CARLO involves two defense approaches. In one approach, the defense processes the predicted results with additional occlusion and distance information extracted from the point cloud data by computing ray castings. The other approach, Sequential View Fusion (SVF), uses an additional front-view segmentation model. 

Both methods were only evaluated with vehicle detection.

\parhead{Evaluation Results and Observations.}
Since CARLO was evaluated using the KITTI dataset\cite{kitti_3d}, we evaluate our removal attack as described in~\S\ref{subsec:kitti_synthesis}. We use the aforementioned ray casting CARLO approach because SVF

requires training a new model for the pedestrian detection task which is not provided in the work by Sun et al.~\cite{sun2020towards}.
 
We set both CARLO threshold rate and PointPillars bounding box IoU threshold to zero.
We then verify that, when the target vehicle/pedestrian is fully removed, the proposed attack achieves a 100\% success rate against both PointPillars and CARLO.
Our results also show that, against CARLO, the minimum attack angle required to remove an obstacle is 2.7$^{\circ}$ and 7.98$^{\circ}$ on average when the target obstacle is a pedestrian and a vehicle respectively. In contrast, the base PointPillars model requires 3.34$^{\circ}$ and 8.77$^{\circ}$ respectively for target pedestrians and vehicles.

Since CARLO was originally designed for detecting spoofing attacks that introduce additional obstacles with a few spoofed points, it introduces false negatives in addition to the base PointPillars model. While measuring our attack results under removing angles that almost fully remove the entire target obstacle, PointPillars can sometimes detect the target obstacle, as discussed in detail by Sun et al.~\cite{sun2020towards}. On the other hand, CARLO identifies such detection as a malicious spoofing attack and removes the spoofed points as a consequence.

\parhead{Hiding Attack Defense.}
Hau~et~al.~\cite{hau2022using} proposed a method against object hiding attacks such as ORA-Random~\cite{hau2021object} by leveraging 3D shadows to locate hidden objects. The defense methodology assumes that the attacker perturbations are limited to short-distance points shifting along the ray direction of the corresponding echoes of the target obstacle to remove. The defense initially identifies shadow regions and detects unlabelled points in the frustums from the detected shadows. We replicate this defense and achieve a 92.9\% object-shadow association rate over the KITTI dataset (benign case).
\parhead{Evaluation Results and Observations.}
We evaluate the defense against our attack using the same methodology as~\S\ref{subsec:kitti_synthesis}. From our 200 samples from the KITTI dataset, we discarded the samples which did not match the required ROI of the defense. Therefore, we use 80 samples (40 pedestrian and 40 vehicles), and observed a 100\% TNR and 0\% TPR when all the cloud points of the target obstacles are removed by our attack. This happens because the defense expects cloud points to exist in the frustum of the detected shadow in order to reveal the presence of an hidden object.

 \subsection{Proposed Defenses}
 \label{subsec:ProposedDefense}
 
 \parhead{Fake Shadow Detection.} We extend the methodology of Hau et al. ~\cite{hau2022using} to include PRA detection. We call this approach Fake Shadow Detection (FSD). The FSD methodology consists first in identifying the shadow regions of the point cloud in the ROI to find potential ORA-based attacks as the previous approach. Then, the shadow regions are compared with the expected shadows of the detected objects from Autoware's euclidean clustering. This comparison is achieved by projecting the point clouds of a detected object on the ground and filtering out the corresponding shadow region based on this projection and the calculated frustum. If the remaining shadow region is above a threshold (\textasciitilde 15 cubic meters based on our empirical experiments on vehicle target obstacles), it is considered a removal attack.

We evaluate our methodology with the 80 PRA samples used previously and additional 80 benign cases and observed 82.5\% TNR and 91.2\% TPR. 
\parhead{Azimuth-based Detection.} 
Despite our upgraded defense FSD methodology reaching high attack detection accuracy, shadow-based approaches remain limited in applicability in real-time AV obstacle detection because of the computational overhead due to projection and frustum operations~\cite{hau2022using}.
Since our attack removes cloud points along a specific direction with respect to the LiDAR, a simpler and practical solution to detect our removal attack is to look for disparities in the raw point cloud data. We can accomplish this by inspecting the horizontal angular view of the LiDAR (azimuth), as our removal attack would create a cloud point gap along the entirety of the attack angle.
Our approach assumes that a LiDAR scan with a gap greater than a 1-degree angle in the azimuth values (minimum attacker capability)  might be a potential removal attack. Thus the approach consists of calculating the azimuth values of all the cloud points in the scene and sorting them in an increasing order based on the calculated values.
Missing cloud points will correspond to missing azimuth regions that, in turn, can reveal the attack.
\parhead{Evaluation Results and Observations.}
To evaluate our approach on benign cases, we run our defense on the KITTI dataset achieving 99.98\% TNR on over 7,480 KITTI scenes.
Then we evaluate the defense against PRA, using the same methodology as in~\S\ref{subsec:kitti_synthesis} over our 200 samples. We synthesize a total of 3,000 scenes with attack angles ranging from 1 - 22 degrees, and we achieve 100\% TPR. The azimuth information also reveals the attack angle and the direction of the attack. 
To evaluate the defense's practicality beyond synthesized attack samples, we also perform the same analysis on our collected traces with the VLP-16 LiDAR. We tested the approach on our 75 benign scenes and 611 attack scenes and observed 100\% TPR and TNR rates, respectively.
\parhead{Runtime Efficiency and Limitations.}
We measure the runtime of the proposed defense on a system with Intel Xeon(R) Silver 4114 CPU (2.2 Ghz x 20) and 64GB RAM. We observed an average runtime of 7.9 ms/scene over the KITTI dataset.
Although this defense shows high performance in PRA detection, the methodology cannot be extended to adversarial spoofing attacks and ORA-based attacks.

%% file: discussion.tex
\section{Discussion}

\parhead{Safety and Ethics Considerations.} 
All experiments were conducted in a controlled environment. Laser safety measures are detailed in the Appendix~\ref{appendix:safety}.
We have notified the LiDAR manufacturer company about our findings.

\parhead{Attack Limitations and Future Work.} One limitation of our attack is that it can only affect a single LiDAR sensor. We do not consider attacking multiple LiDARs, solid-state LiDARs, or groups of LiDARs and cameras. We demonstrate the attack success over LiDAR-camera fusion models by only attacking the spinning LiDAR. This could be because AV perception models mostly relies on such sensors for obstacle detection rather than cameras as indicated by Cao et.al~\cite{cao2019adversarial}. A future analysis might include testing on different models and AD frameworks. The considerations on the generality of the proposed attack methodology are described in Appendix~\ref{appendix:generality}.
\parhead{Engineering Limitations.}
Our setup has limitations for attacking a moving vehicle. Though our proof-of-concept experiments in section~\S\ref{subsec:tracking} demonstrate the feasibility of the proposed attack using basic tracking equipment, there are several engineering limitations for further extending the attack success rate for moving AVs. First, attacking the LiDAR at a distance is non-trivial. The LiDAR scan is sparser at a distance which affects the synchronization with the LiDAR. Therefore, larger optical devices or more photodiodes are required for synchronization. The spoofed signals also diverge at a distance, which affects the intensity of the spoofed signals. Therefore, a more sophisticated optical system for converging the spoofed signals is required. Second, aiming at the LiDAR requires a good object detection model. Since the LiDAR sensor models can vary on the AVs, a general model for different LiDAR sensors in different environments is necessary for the attack in the real world. Lastly, aiming the LiDAR moving at a higher speed requires higher precision. The attacker needs to either use a robotic arm with higher resolutions of movements or use a laser diode with higher power such that it can attack a larger area at the same time.

%% file: conclusion.tex
\section{Conclusion}

We discover a new physical removal attack which removes LiDAR point cloud from genuine obstacles. This study explores the attacker capability to use the point cloud ablation at the sensor level to cause the AD perception module to fail to recognize obstacles and their locations, reaching an attacker capability of $45^\circ$. We then evaluate the effect of PRA on three AV perception and fusion models. We also achieve a $92.7\%$ success rate removing 90\% of a target obstacle cloud point on a moving vehicle. Finally, we propose two effective defense strategies to help mitigating the threat.

%% file: appendix.tex
\appendix
\section*{Appendices}

\section{Laser Safety Details}
\label{appendix:safety}
During the outdoor experiments, at most 4,000 pulses per scan (100 ms) of the LiDAR are emitted by the spoofer. According to the datasheet of the laser-diodes\cite{osram_diode}, the typical power during emission is around 70 W. Since the spoofing device emits pulses with 40 ns width we can first calculate the per pulse energy as below:
\begin{equation*}
     E = 70\,W \cdot 40\,ns = 2.8\,\mu J 
\end{equation*}

By calculating the maximum permissible exposure (MPE) of the laser pulses at 905 $nm$ wavelength:
\begin{equation*}
    MPE = 18 \cdot t^{0.75} \cdot 10 ^ {(905-700)/500}
\end{equation*}
when t is 0.25 s, which is the time taken by a blink of the eye, the estimated MPE is 6.36 $J/m^2$. Since the total number of pulses during that period is less than $4000*3=12000$, we can calculate the minimum radiated area as:
\begin{equation*}
    A = 2.8 \,\mu J / 6.36 \,J/m^2  = 26.42 \,mm^2  
\end{equation*}
where the diameter of the radiated area in our setup is around 11.6 $mm$. This diameter is smaller to the measured size which proves the safety for outdoor experiments.

\section{Generality of the Methodology}
\label{appendix:generality}

Our PRA exploit the automatic transformation and filtering process common to many commercial spinning LiDARs and popular AD frameworks. We demonstrate how three different object detection models, and three fusion models for AVs are susceptible to the attack and demonstrate the consequences on a industry-grade AD simulator system. Thus our attack approach can be generalized to other LiDAR-based AV perception systems. We also observe how different LiDAR sensors adopt the same interface and filter phases of the Velodyne VLP-16 (see Table~\ref{table:min}). Thus the transformation chains of other sensors are likely to also be vulnerable to such physical attacks. In addition, as demonstrated in previous works~\cite{cao2019adversarial,sun2020towards}, the LiDAR sensor spoofing capability can also be generalized because it is independent from the AV system.